\documentclass[11pt]{article}

\usepackage{hyperref}
\usepackage{amsthm}
\usepackage{graphicx, color, url}
\usepackage{amsmath}
\usepackage{dsfont}
\usepackage{amssymb}
\usepackage{epstopdf}
\usepackage{boxedminipage}
\usepackage{amsfonts}
\usepackage{url}
\usepackage{amsmath}
\usepackage{amsbsy}
\usepackage{setspace}
\usepackage{natbib}
\usepackage{bm}
\usepackage{wasysym}
\usepackage{textcomp}
\usepackage{bbm}
\usepackage{verbatim}
\usepackage{enumitem}
\usepackage[ruled, noend]{algorithm2e} 

\newtheorem{theorem}{Theorem}
\newtheorem{corollary}{Corollary}
\newtheorem{proposition}{Proposition}
\newtheorem{lemma}{Lemma}
\newtheorem{example}{Example}
\newtheorem{remark}{Remark}
\newtheorem{definition}{Definition}

\newcommand{\beq}{\begin{equation}}
	\newcommand{\eeq}{\end{equation}}
\newcommand{\beas}{\begin{eqnarray*}}
	\newcommand{\eeas}{\end{eqnarray*}}
\newcommand{\bea}{\begin{eqnarray}}
	\newcommand{\eea}{\end{eqnarray}}
\newcommand{\bei}{\begin{itemize}}
	\newcommand{\eei}{\end{itemize}}
\newcommand{\ben}{\begin{enumerate}}
	\newcommand{\een}{\end{enumerate}}
\newcommand{\bet}{\begin{theorem}}
	\newcommand{\eet}{\end{theorem}}
\newcommand{\bel}{\begin{lemma}}
	\newcommand{\eel}{\end{lemma}}
\newcommand{\bep}{\begin{proposition}}
	\newcommand{\eep}{\end{proposition}}
\newcommand{\bed}{\begin{definition}}
	\newcommand{\eed}{\end{definition}}
\newcommand{\bec}{\begin{corollary}}
	\newcommand{\eec}{\end{corollary}}
\newcommand{\bex}{\begin{example}}
	\newcommand{\eex}{\end{example}}

\newcommand{\II}{\mathbb I}
\newcommand{\EE}{\mathbb E}

\newcommand{\argmax}{\mathop{\rm arg\max}}

\newcommand{\hlfdr}{\widehat{\mbox{Clfdr}}}
\newcommand{\hQ}{\hat{Q}}
\newcommand{\bdelta}{\boldsymbol{\delta}}

\SetKwInput{KwInput}{Input}                
\SetKwInput{KwOutput}{Output} 

\makeatletter
\newcommand{\customlabel}[2]{%
	\protected@write \@auxout {}{\string \newlabel {#1}{{#2}{\thepage}{#2}{#1}{}} }%
	\hypertarget{#1}{#2}
}
\makeatother

\begin{document}
	
	
 	\title{Ranking and Selection in Large-Scale Inference of Heteroscedastic Units} 
  	\author{Bowen Gang$^1$, Luella Fu$^2$, Gareth James$^3$ ,and Wenguang Sun$^4$}
	\date{}
	
	\maketitle
 \begin{abstract}
    
The allocation of limited resources to a large number of potential candidates presents a pervasive challenge. In the context of ranking and selecting top candidates from heteroscedastic units, conventional methods often result in over-representations of subpopulations, and this issue is further exacerbated in large-scale settings where thousands of candidates are considered simultaneously. 
To address this challenge, we propose a new multiple comparison framework that incorporates a modified power notion to prioritize the selection of important effects and employs a novel ranking metric to assess the relative importance of units. We develop both oracle and data-driven algorithms, and demonstrate their effectiveness in controlling the error rates and achieving optimality.  We evaluate the numerical performance of our proposed method using simulated and real data. The results show that our framework enables a more balanced selection of effects that are both statistically significant and practically important, and results in an objective and relevant ranking scheme that is well-suited to practical scenarios. 
\end{abstract}

		\vspace{9pt}
 
	\footnotetext[1]{Department of Statistics and Data Science, Fudan University, Shanghai, 200437, China, bgang@fudan.edu.cn.}  
	\footnotetext[2]{Department of Mathematics, San Francisco State University, San Francisco, CA 94132, U.S.A., luella@sfsu.edu.}  
	\footnotetext[3]{Department of Information Systems and Operations Management, Emory University, Atlanta, GA 30322, U.S.A., gareth@emory.edu. } 
 	\footnotetext[4]{School of Management and Center for Data Science, Zhejiang University, Zhejiang, 310058, China, wgsun@zju.edu.cn. } 
	
		\thispagestyle{empty}

	\newpage
	\renewcommand{\theequation}{\thesection.\arabic{equation}}

	\setcounter{section}{0} 
	\setcounter{equation}{0} 

	\setcounter{page}{1} 
	

	\thispagestyle{empty}

	\newpage
	\renewcommand{\theequation}{\thesection.\arabic{equation}}

	\setcounter{section}{0} 
	\setcounter{equation}{0} 

	\setcounter{page}{1} 


\section{Introduction}

Allocating limited resources among numerous potential candidates is a common problem faced by both individuals and organizations. This dilemma is encountered by NBA basketball recruiters as they search for promising talents, public policy makers as they fund educational programs, and  internet users on platforms such as Yelp as they decide which restaurants to visit. Such decision-making scenarios give rise to the ranking and selection problem, a fundamental statistical issue that requires the comparison of multiple unknown parameters.

Ranking and selection  has been a classical topic in multiple comparisons (\citealp{1948Mostellar, 1949Paulson, 1954Bechofer, 1965Gupta, Panchapakesan1971, GoelRubin1977}), and its integration into other branches of statistics, operations research, and computing has made it a critical and constantly evolving area of study (\citealp{Chenetal2000, 2012Boydetal, Luoetal2015, Nietal2017, KaminskiPrzemyslaw2018,  Zhongetal2022}). The decision process has two critical components: first, establishing a meaningful criterion for ordering a pool of potential candidates, and second, selecting a subset of ``most meritorious'' candidates with a certain level of confidence. Properly accounting for the heteroscedasticity across data from diverse study units is essential for producing effective, sensible, and fair decisions in the ranking and selection process.  
In what follows, we first provide an overview of conventional practices and identify relevant issues, followed by an exposition of our new framework for addressing the challenge of heteroscedasticity. Finally, we discuss related works and highlight the contributions of our approach.

\subsection{Conventional practices and issues}

Ranking is essential in multiple comparisons to evaluate and identify top-performers from a pool of potential candidates. While the importance of each candidate is linked to the magnitude of its associated parameter, the decision-making process also takes into account the associated uncertainties in order to ensure that the top candidates indeed belong to the ``most meritorious'' group. The two perspectives, namely the parameter magnitude and the confidence level in the assertions being made, are reflected by the estimated effect size and its associated statistical significance, respectively. In homoscedastic models, these two perspectives yield the same ranking. However, in cases where the data are heteroscedastic across the study units, the rankings based on these two perspectives may disagree. As demonstrated shortly, the issue is further exacerbated  in large-scale settings where thousands of candidates are being considered at once. Developing a sensible ranking and selection criterion that partially mitigates the conflict between the two perspectives poses a critical challenge in large-scale multiple comparison problems. 

\begin{figure}
	\includegraphics[width=6in]{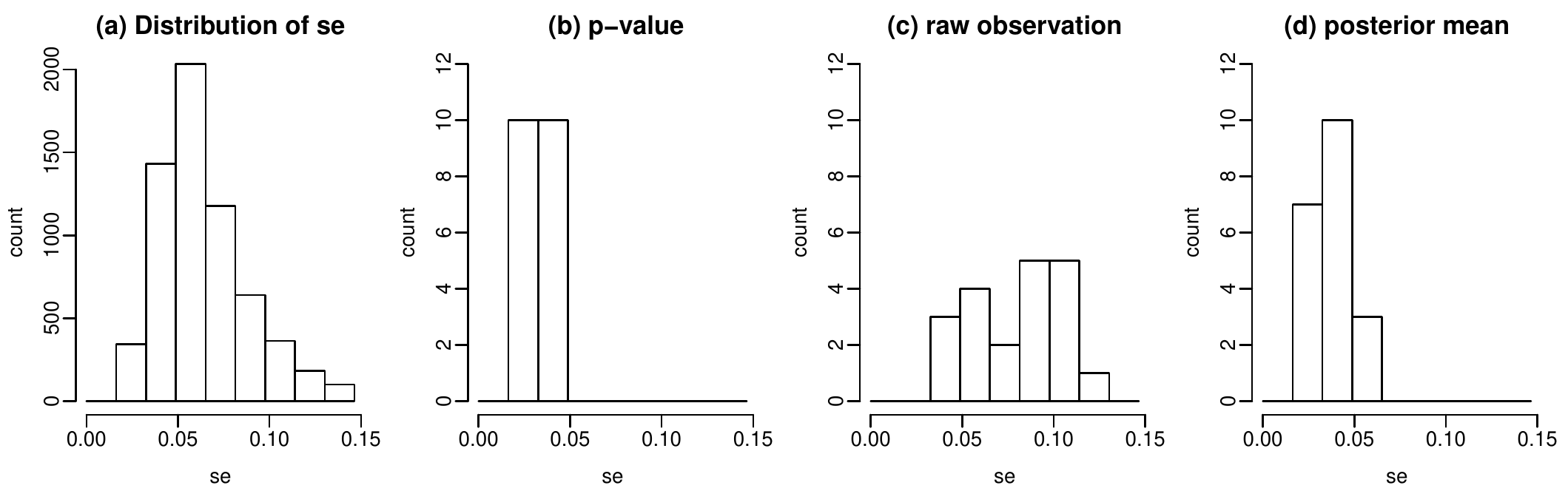}
	\caption{Panel (a): overall distribution of SEs of the AYP data set. Panel (b): distribution of SEs of the top 20 schools according to p-value. Panel (c): distribution of SEs of the top 20 schools according to raw observation. Panel (d): distribution of SEs of the top 20 schools according to posterior mean.
	}
	\label{badselection}
\end{figure}

To demonstrate the inadequacy of conventional practices, we analyze the 2005 Annual Yearly Performance (AYP) data to identify K-12 schools with significant gaps in passing rates between socioeconomically advantaged (SEA) and disadvantaged (SED) students. The raw observations are the empirical differences in passing rates between the two groups, with the standard errors (SEs) being  linked to the number of students in the schools. More details of the study are provided in Section \ref{sec:ayp}. We consider three selection strategies, which are respectively based on statistical significance ($p$-value), observed gap in passing rates (raw observation), and posterior mean (computed using Tweedie's formula). The results of our exploratory  analysis are presented in Figure \ref{badselection}. Panel (a) shows the distribution of the SE. Panels (b), (c) and (d) show the distribution of 20 selected schools according to $p$-value, observed gap and posterior mean, respectively. Figure \ref{badselection} reveals that selecting schools based on $p$-values and posterior means tends to result in an over-representation of schools with low SEs, while selecting based on raw observations may lead to an over-representation of those with high SEs. The design of this analysis draws on earlier works, including \cite{SunMcLain:2012} and \cite{HendersonNewton2014}, which have identified and provided initial insights into some perplexing phenomena that arise under heteroscedastic models. For example, \cite{SunMcLain:2012} found that the largest 1\% of K-12 schools are over-represented among the worst performing ten schools when the selection is based on p-values. 

Although all three selection criteria have their advantages in capturing either the effect sizes or accounting for associated uncertainties, the over-representation of subgroups with high/low SEs is undesirable and runs counter to practical wisdom. We aim to develop a new ranking and selection framework that resides between the three polarized selection criteria, striking a balance between their respective advantages and disadvantages. The AYP data will be revisited under our new framework in Section \ref{sec:ayp}. 

\subsection{False discovery rate analysis under heteroscedasticity}

We begin by examining the use of the false discovery rate (FDR) framework \citep{bh1995, Storey2002, GenoveseWasserman2004} in the context of selecting important candidates. Most FDR methods operate in two steps: ranking and thresholding, where the building block of operation is the $p$-value \citep{bh1995} or the local false discovery rate (lfdr; \citealp{efron2001empirical, sun2007oracle}). Both the $p$-value and lfdr tend to prioritize the stability of data over effect sizes, which leads to the over-selection of schools with low SEs in the AYP analysis. To comprehend the limitations of the conventional formulation, we can refer to the theory in \cite{sun2007oracle}, which shows that thresholding lfdr is optimal in the sense that it maximizes the average power subject to the constraint on the FDR. This perspective reveals two major issues that contribute to the difficulties of utilizing the FDR framework in heteroscedastic models.

The first issue is that the concept of average power, which is defined as the expected proportion of non-nulls that are correctly rejected, overlooks the severity of missed signals. This gives rise to a significant limitation that is particularly concerning in situations with substantial heteroscedasticity across units. Specifically, the identification of a large signal should be rewarded more than that of a small signal, even if both study units have the same level of statistical significance. However, this principle is not fulfilled by the conventional FDR formulation. To correct the inherent bias in conventional FDR analyses, it is desirable to modify the power concept such that selection of larger effects is prioritized with a higher reward. 

The second issue pertains to the conventional multiple testing framework, which employs a thresholding procedure that is contingent on a fixed ordering determined by a predefined significance index, such as the $p$-value or lfdr. However, our optimality theory on prioritized selection reveals that the existence of such an ordering is not guaranteed. The absence of a universally optimal ordering poses a significant challenge in developing an objective ranking, as the rankings can be inconsistent across users who may judiciously select different confidence or reference levels.

\subsection{A preview of the proposed method}

Our proposal presents a new multiple comparison framework that addresses the two aforementioned issues by incorporating a modified power notion to prioritize the selection of important effects and employing a novel ranking index to assess the relative importance of units. 

We first study the prioritized selection problem by utilizing a constrained optimization formulation. The goal is to control a user-specified FDR while maximizing a modified power concept that assigns higher rewards to selections of larger effects. The solution leads to a selection method that carefully weighs the candidate's effect size against its significance. The new formulation reduces the bias inherent in commonly used significance indices that favor stability, ensuring that the effect size is more fairly represented in the selection process.

We then turn to the ranking issue by introducing a novel concept called the ``r-value,'' which provides a  measure of the relative importance of study units in a list. The importance of different units is captured by how early they are selected according to a varying target. The earlier a unit is selected, the more important it is considered to be relative to the other units -- thus an objective ranking of study units is generated.

\subsection{Our contributions}

In scenarios where study units display substantial heteroscedasticity, the proposed ranking and selection procedure offers a valuable alternative to conventional FDR analyses. Our method enables a more balanced selection of effects that are both statistically significant and practically important, resulting in a ranking that is objective and relevant for practical scenarios. To tackle the complexity that arises from our revised notion of power, we have devised an oracle procedure and developed a theory to establish its optimality. The new theory offers a significant advance in contrast to the weighted FDR theory in \cite{Basuetal2018}. Furthermore, we have developed a computational shortcut of the oracle procedure, and rigorously established the asymptotic properties of the corresponding data-driven algorithm. Our work presents a unified framework that explicitly incorporates considerations of effect size, statistical significance, error control, theoretical guarantees, and computational efficiency for analyzing heteroscedastic data.

Previous studies have made progress in addressing some, but not all, of our challenges. \cite{SunMcLain:2012} proposed a decision-theoretic framework that incorporates information about effect sizes, but their approach relies on standardization and does not resolve the issue of over-representation of small variances. \cite{HendersonNewton2014} put forth the maximal agreement method to avoid over-representation. However, their formulation differs significantly from ours in two aspects: firstly, the question of error rate control is left unaddressed, and secondly, the joint consideration of effect size and significance is absent. \cite{GuKoenker2020invidious} devised a robust set of ranking and selection methods within a compound decision-theoretic framework. Notably, they extended the maximal agreement method to include false discovery control. However, the challenge of balancing statistical significance and effect size has not been fully resolved. Finally, \cite{fu202hart} demonstrated that standardization can distort structural information about the alternative distribution, but their analysis had focused on the conventional FDR framework.

\subsection{Organization}

The paper is structured as follows. 
Section \ref{sec:method} presents the problem formulation and an oracle procedure for prioritized selection. In Section \ref{sec:dd}, we develop a data-driven procedure and establish its theoretical guarantees. Section \ref{sec:r-value} introduces the r-value and discusses its agreeability property. 
Sections \ref{sec:numerical} and \ref{data} present results to illustrate the numerical performance of our proposed ranking and selection methods by using both simulated and real data.

\section{Prioritized Selection with FDR Control}\label{sec:method}

This section first introduces the model, notation and problem formulation, then proposes an oracle procedure for prioritized selection of important effects. 

\subsection{Problem formulation}
Suppose $X_i$, $i\in [m]\equiv \{1, \cdots, m\}$ are independent observations from a random mixture model
with possibly heteroscedastic errors:
\begin{equation}\label{data.assump1}
	X_i=\mu_i+\epsilon_i, \;\;\; \epsilon_i\;{\sim}\; N(0,
	\sigma_i^2),
\end{equation}

where $\mu_i$ and $\sigma_i$ come from unspecified distributions with bounded supports:
\begin{equation}\label{data.assump2}
	\mu_i \sim g_\mu(\cdot) ,  \qquad \sigma_i^2 \sim g_{\sigma}(\cdot), \quad   i\in[m].
\end{equation}

To focus on the central idea, we assume that $\sigma_i$ are known, a common practice pursued, for example, in \citet{Efron:2011:Tweedie}, \citet{Xieetal2012}, and \citet{Weinsteinetal18}. The issue of estimating unknown and heterogeneous $\sigma_i$ has been considered in \citet{GuKoenker2017empirical}, \citet{GuKoenker2017unobserved}, \citet{Banetal21-NEST}, \citet{kwon2023f}.

Let $\mathcal A_i$ be a user-specified indifference region. Without loss of generality, suppose one wishes to test whether the effect size $\mu_i$ surpasses a given threshold $\mu_0$, hence $\mathcal{A}_i=\{\mu:\mu\leq \mu_{0}\}$. Upon observing $(x_i, \sigma_i)$, the null and alternative hypotheses are

\begin{equation}\label{data.hypoths}
	H_{0, i}: \mu_i \in \mathcal{A}_i \quad \text{vs.} \quad  H_{1, i}: \mu_i \notin \mathcal{A}_i. 
\end{equation}
Denote $\theta_i=\II(\mu_i>\mu_0)$ the true state of the $i$th item, and $\delta_i\in\{0, 1\}$ the decision we make about that item, where $\delta_i = 1$ if the $i$th item is selected (or claimed as an important case) and $\delta_i=0$ otherwise. Let $\pmb\delta=(\delta_1, \ldots, \delta_m)$. 

In large-scale selection problems, 
a practical and effective goal is to control the false discovery rate \citep{bh1995}
	$$
	\text{FDR}(\bdelta) = \EE\left[\frac{\sum_{i}(1-\theta_i)\delta_i}{(\sum_{i}\delta_i) \vee 1}\right],
	$$	
	where $a\vee b=\max(a,b)$. A closely related quantity is the  marginal false discovery rate 
	$$
	\mbox{mFDR}(\bdelta)=\frac{\EE\left(\sum_i (1-\theta_i)\delta_i\right)}{\EE\left(\sum_i \delta_i \vee 1 \right) }.
	$$

	Under certain first- and second-order conditions, the mFDR asymptotically equals the FDR \citep{genovese2002operating,CaiSunWang2019}. For theoretical convenience we adopt mFDR in our discussion.

	In conventional FDR analysis, the goal is to find a decision rule $\pmb\delta$ that controls the error rate at pre-specified level $\alpha$ with the largest power. A widely used metric for evaluating the power of a multiple testing procedure is the expected number of true positives	
	\beq \label{eq:ETP}
	\mbox{ETP}(\bdelta)=\EE\left(\sum_i \theta_i \delta_i\right)=\EE\left\{\sum_i \II(\mu_i>\mu_0)\delta_i\right\}.
	\eeq

	To prioritize the selection of large effects, we propose to modify the power concept as 
	\begin{equation}\label{eq:ETP-star}
		\mbox{ETP}^*({\bdelta})=\EE\left\{\textstyle\sum_i(X_i - \mu_0)\delta_i\right\}. 
	\end{equation}
	The traditional power concept \eqref{eq:ETP} has undergone two modifications, initially replacing the indicator $\II(\mu_i-\mu_0)$ with the actual difference $(\mu_i-\mu_0)$, followed by substituting the unbiased estimate $X_i$ in place of $\mu_i$, resulting in the revised power concept \eqref{eq:ETP-star}. The first modification allows for the revised power metric to precisely capture the impact of signal magnitude, while the subsequent alteration is crucial for avoiding technical intricacies, as the unknown $\mu_i$ poses a significant difficulty in constructing the oracle rule in Section \ref{oracle}. 
	
	The above considerations give rise to the following constrained optimization problem, in which we aim to develop a selection rule $\pmb\delta\in\{0, 1\}^m$ to
	\beq\label{eq:RS-problem2}
	\mbox{maximize  $\mbox{ETP}^*({\bdelta})$\quad  subject to\quad  $\mbox{mFDR}(\bdelta)\leq\alpha$}. 
	\eeq
	
	\begin{remark}	\rm{
			Our formulation can be extended by replacing $(X_i-\mu_0)$ with a more general function $h(X_i)$. If one only cares about detecting the true state of nature and ignores the severity of missed signals, then we can take $h(X_i)=\II(X_i>\mu_0)$. The other possible choice for $h(X_i)$ is $(X_i - \mu_0)_+$, which ensures that the weight is always positive. Moreover, the choice of $h(X_i)=(X_i - \mu_0)_+$ simplifies subsequent analyses. However, our preference lies with $(X_i-\mu_0)$ over $(X_i - \mu_0)_+$, as the former penalizes the identification of small effects. This preference is in line with the objective of our formulation, which aims to allocate a more balanced representation to the effect size during the selection process. In Section~\ref{oracle}, we will demonstrate the critical role of the sign of $h(X_i)$ in the analysis. In contrast to the weighted FDR problem discussed in \cite{Basuetal2018}, where the weights $w_i$ are assumed to be non-negative and independent of $X_i$, the ``weights'' $h(X_i)$ in our formulation are allowed to be negative and depend on $X_i$. The difference poses new challenges in developing both oracle and data-driven procedures. We discuss related issues in subsequent sections.
		}
	\end{remark}	
	
	\subsection{Oracle selection procedure}\label{oracle}
	
	This section considers an ideal scenario where an oracle knows $g_\mu$ and $g_\sigma$ in \eqref{data.assump2}. The oracle rule weighs the tradeoffs between $\alpha$-investing and $\mu$-investing processes, two concepts that we shall elaborate on shortly, and assesses their impacts on the modified power and FDR capacity, respectively. In what follows, we present a heuristic argument to explain how we arrive  at the oracle rule, and rigorously prove its optimality in Theorem \ref{thm:or}. 
	
	The process of $\alpha$-investing \citep{foster2008alpha, gang2021structure}, which is used to evaluate the gains and losses in making a discovery, relies on the conditional local false discovery rate (Clfdr, \cite{sun2009large, efron2012large, SunMcLain:2012}). The Clfdr is defined as
	\beq\label{lfdr-st}
	\mbox{Clfdr}_i=\mathbb P(\mu_i \in \mathcal{A}_i|x_i,\sigma_i)=\dfrac{f_{0i}(x_i)}{f_i(x_i)},
	\eeq
	where $f_{0i}(x_i)=\int_{\mu \in \mathcal{A}_i}^{}\phi_{\sigma_i}(x_i-\mu)g_\mu(\mu)d\mu$ and $f_i(x_i)=\int_{-\infty}^{\infty}\phi_{\sigma_i}(x_i-\mu)g_\mu(\mu)d\mu$.
	The ordered values of Clfdr are denoted $\text{Clfdr}_{(1)},\ldots,\text{Clfdr}_{(m)}$. As shown by \citet{sun2007oracle}, the following step-wise algorithm, which uses the Clfdr as a basic operation unit, is asymptotically optimal in the sense that it maximizes the ETP subject to the constraint $\mbox{mFDR}\leq \alpha$. 
	\beq\label{AZ}
	\mbox{Let $k=\max\left\{j: \frac 1 j\sum_{i=1}^j\text{Clfdr}_{(i)}\leq \alpha\right\}$, then reject $H_{(1)}, \ldots, H_{(k)}$}.
	\eeq
	The Clfdr algorithm \eqref{AZ} can be interpreted as a varying-capacity knapsack process \citep{Basuetal2018,gang2021structure}. Specifically, \eqref{AZ} can be viewed as an iterative decision process where the initial $\alpha$-wealth is invested by rejecting hypotheses sequentially. The process adheres to the following constraint: 
	\begin{equation}\label{FDR-cons} 
		\mbox{Clfdr}_j - \alpha \leq C_j \equiv - \sum_{H_i\in \mathcal R_{j}} \left(\mbox{Clfdr}_i - \alpha\right), \mbox{ for $j=1, 2, \ldots$, }
	\end{equation}
	where $\mathcal R_j\subset \{H_1, H_2, \ldots, H_m\}$ is the collection of rejected hypotheses at step $j$, 
	and $C_j$ may be viewed as the \emph{capacity} of the knapsack at step $j$, with the default choice $C_1=0$. 
	Under this view, the $\alpha$-investing process corresponds to a knapsack problem whose capacity may either expand or shrink over time. If  $H_j$ with $\mbox{Clfdr}_j<\alpha$ ($\mbox{Clfdr}_j>\alpha$) is rejected, then $C_j$ increases (decreases) by $|\alpha-\mbox{Clfdr}_j|$. 
	
	The $\mu$-investing process, on the other hand, is relatively straightforward. When a hypothesis $H_j$ with $X_j>\mu_0$ ($X_j<\mu_0$) is rejected, the return on investment increases (decreases) empirically by $|X_j-\mu_0|$. 
	
	Jointly considering the gains and losses in the $\alpha$-investing and $\mu$-investing processes, we divide the hypotheses into four groups:
	\begin{enumerate} 
		\setcounter{enumi}{-1}
		\item $X_i-\mu_0\geq 0$ and $\mbox{Clfdr}_i-\alpha\leq 0$;
		\item $X_i-\mu_0\geq 0$ and $\mbox{Clfdr}_i-\alpha> 0$; 
		\item $X_i-\mu_0<0$ and $\mbox{Clfdr}_i-\alpha\leq0$;
		\item $X_i-\mu_0<0$ and $\mbox{Clfdr}_i-\alpha>0$.
	\end{enumerate}
	Our problem formulation suggests that units in group 0 should always be selected, as their selection results in an increase in both $\alpha$-wealth and power. Conversely, units in group 3 should never be selected, as their selection leads to decreases in both $\alpha$-wealth and power. The tradeoffs involved in selecting units from groups 1 and 2 are nuanced. In the case of group 1, selecting units comes at the cost of sacrificing capacity, but also results in increased power. We hypothesize that the optimal strategy involves selecting units with a high value-to-cost ratio, defined as 
	\begin{equation}\label{T-OR}
		T_i = \frac{X_i - \mu_0}{\mbox{Clfdr}_i - \alpha}, 
	\end{equation}
	By contrast, selecting units from group 2 involves trading power for increased capacity. Consequently, the $T_i$ statistic can be viewed as a cost-to-value ratio. Therefore, it is desirable to select units with low values of $T_i$ from group 2. For theoretical convenience, we consider a bounded, continuous and monotone transformation $\xi(\cdot)$ of $T_i$. Let $S_i = \xi(T_i)$. An example of such a transformation could be the hyperbolic tangent function $\xi(x) \equiv \tanh(x) = (e^x - e^{-x}) / (e^x + e^{-x})$. We hypothesize that the optimal decision rule can be expressed in the following form:
	\begin{equation}\label{opt-rule}
		\delta(c_1,c_2)(S_i) = 
		\begin{cases}
			1 & \mbox{if } \ (X_i,\text{Clfdr}_i) \ \mbox{belongs to group 0} \\
			1 &\mbox{if } \ (X_i,\text{Clfdr}_i) \ \mbox{belongs to group 1 and }S_i>c_1 \\
			1 &\mbox{if } \ (X_i,\text{Clfdr}_i) \ \mbox{belongs to group 2 and }S_i< c_2 \\
			0 &\mbox{otherwise}
		\end{cases},
	\end{equation}
	where $c_1$ and $c_2$ are thresholds to be determined. 
	Define 
	$$
	c^-_1=\inf_{(X,\text{Clfdr}) \in \mbox{group 1}}\xi(T),\quad 	c^+_1=\sup_{(X,\text{Clfdr}) \in \mbox{group 1}}\xi(T),$$
	$$ 	c^-_2=\inf_{(X,\text{Clfdr}) \in \mbox{group 2}}\xi(T) ,\quad c^+_2=\sup_{(X,\text{Clfdr}) \in \mbox{group 2}}\xi(T).
	$$
	We consider a class of decision rules of the form \eqref{opt-rule}, and denotes its mFDR and modified power by $\mbox{mFDR}(c_1, c_2)$ and $\mbox{ETP}^*(c_1,c_2)$ respectively. 
	Note that $\mbox{mFDR}{(c_1,c_2)}$ and $ \mbox{ETP}^*{(c_1,c_2)}$ are both continuous and bounded functions of $(c_1,c_2)$. Moreover, both are constant outside of the rectangle $\left[  c^-_1 ,c^+_1\right] \times \left[ c^-_2 , c^+_2\right] $. Hence, without loss of generality, we restrict $\mbox{mFDR}{(c_1,c_2)}$ and $\mbox{ETP}^*{\pmb{\delta}(c_1,c_2)}$ to the compact set $\left[  c^-_1 ,c^+_1\right] \times \left[ c^-_2 , c^+_2\right] $. 
	Define 
	\begin{equation}\label{opt-cutoffs}
		(c^{OR}_1,c^{OR}_2)=\argmax_{(c_1,c_2)\in \left[  c^-_1 ,c^+_1\right] \times \left[ c^-_2 , c^+_2\right] }\{ \mbox{ETP}^*{(c_1,c_2)}: \mbox{mFDR}{(c_1,c_2)}=\alpha \}.
	\end{equation}
	We state our first main result:
	\begin{theorem}\label{thm:or}
		The oracle procedure $\pmb{\delta}^{OR}=\pmb{\delta}(c^{OR}_1,c^{OR}_2)$ proposed above controls mFDR at level $\alpha$ and is optimal in the sense that for any decision rule $\pmb{\delta}$ that controls mFDR at level $\alpha$, {we always have $ \mbox{ETP}^*{(c_1^{OR},c_2^{OR})} \geq  \mbox{ETP}^*_{\pmb{\delta}}$}. 
	\end{theorem}

	\subsection{Incorporating effect size: an illustrative example}

	This section provides a toy example to contrast two oracle rules designed to maximize the conventional power \eqref{eq:ETP} and modified power \eqref{eq:ETP-star}, respectively. The fundamental operational units for the two oracle rules are $\mbox{Clfdr}$ and $T$, defined by \eqref{lfdr-st} and \eqref{T-OR}, respectively.  
	
	Suppose we are interested in testing $H_{0,i}:\mu_i\leq 0$, $i\in[m]$ based on data generated from the following model: 
	$$
	\theta_i\overset{i.i.d.}{\sim}\text{Bernoulli}(0.2),\quad \mu_i\overset{ind}{\sim}(1-\theta_i)U(-3,-1)+\theta_iU(1,2),\quad \sigma_i\overset{iid}{\sim }U(0.5,3),\quad X_i\overset{ind}{\sim} \mathcal N(\mu_i,\sigma^2_i).
	$$
	
	The oracle rule $\pmb\delta^C$ that maximizes the ETP in equation \eqref{eq:ETP} is defined as $\pmb\delta^C=(\delta^C_i:i\in[m])$, where $\delta_i^C=\II (\mbox{Clfdr}_i<c_\alpha)$ and $c_\alpha$ is determined by the desired FDR level $\alpha$. For the oracle rule $\pmb\delta^{T}=(\delta^{T}_i:i\in[m])$ maximizing $\mbox{ETP}^*$ \eqref{eq:ETP-star}, Group 2 is empty, and the oracle rule for units in Group 1 is $\delta_i^{T}=\II (T_i>t_\alpha)$. Assuming known distributional information, we can determine $c_\alpha=0.32$ and $t_\alpha=12.21$ through numerical approximations such that the FDR levels of both oracle rules are exactly controlled at $\alpha=0.1$. 
	
\begin{figure}
		\includegraphics[width=6.5in]{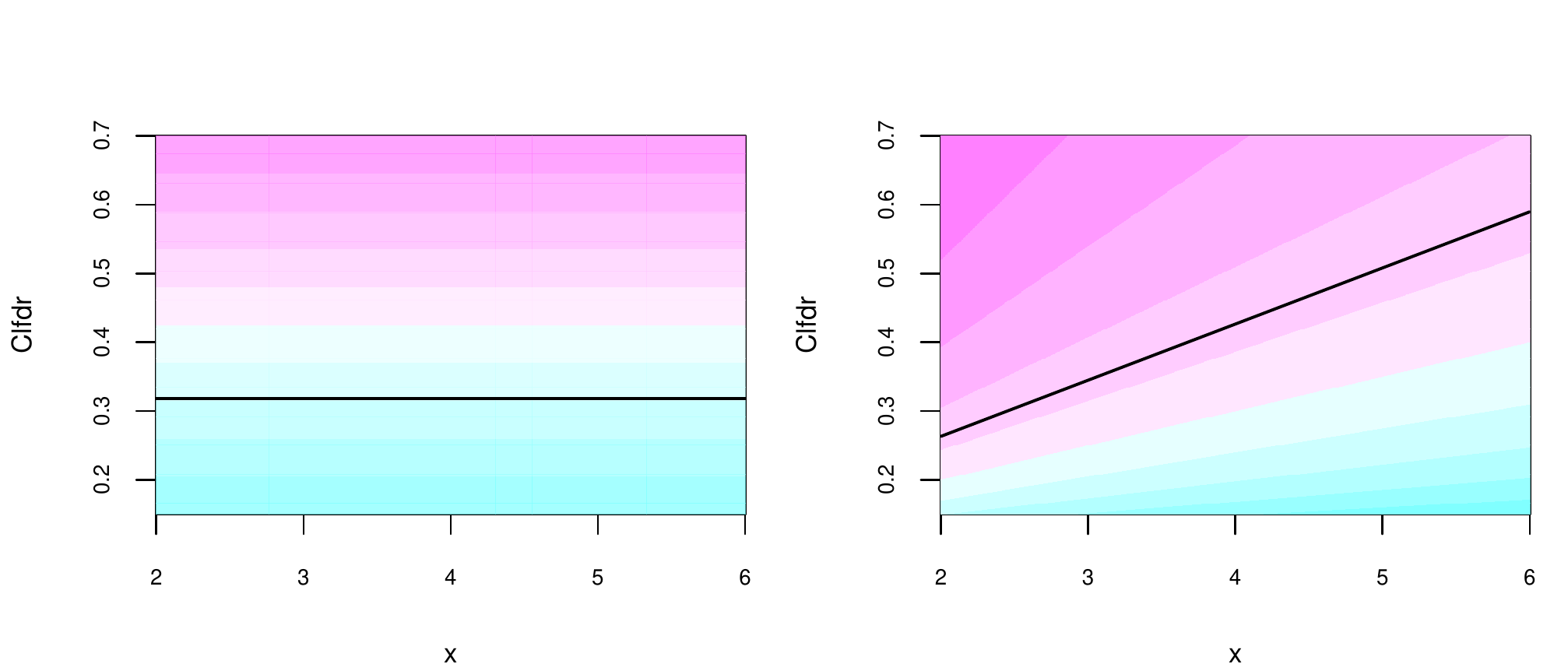}
		\caption{ Left: heat map for Clfdr values. Right: heat map for T values. In both panels, the x-axis and y-axis represent raw observations $X$ and Clfdr values, respectively. The rejection regions of the oracle rules $\pmb\delta^{C}$ and $\pmb\delta^{T}$ are the areas under the corresponding black lines. }
		\label{illuseg}
	\end{figure}

	Figure \ref{illuseg} displays the rejection regions of the oracle rules $\pmb\delta^{C}$ and $\pmb\delta^{T}$, depicted by the corresponding black lines. The left panel of Figure \ref{illuseg} shows the heat map for Clfdr values, with the x-axis representing raw observations $X$ and the y-axis representing Clfdr values. On the right panel, we present the heat map for $T$ values, with the x-axis and y-axis representing $X$ and  Clfdr values, respectively. We observe that, in contrast to the patterns in the left panel, where the rejection region only depends on Clfdr values, our new oracle rule yields a rejection region, shown on the right panel, where the Clfdr threshold for the oracle increases as $X$ increases. It is clear that our oracle rule with prioritized selection depends on both statistical significance and effect size, and exhibits a preference for selecting units with larger $X$ compared to the oracle rule that aims to maximize conventional power. 
	
	\section{Data-driven procedure }\label{sec:dd}
	
	This section presents the development of our data-driven procedure, including a non-parametric deconvoluting method (Section \ref{deconv}), a computational shortcut (Section \ref{shortcut}), and a theoretical analysis  (Section \ref{theory}).
	
	
	\subsection{Nonparametric deconvolution}\label{deconv}

	We propose a non-parametric $g$-modeling approach to estimating $g_\mu(\cdot)$, which plays a critical role in computing the value of $S_i$. Although prior research by \cite{efron2016deconv}, \citet{GuKoenker2017unobserved} and \cite{GuKoenker2020invidious} has tackled this issue, the theoretical properties of these methods remain largely unknown. Our new $g$-modeling method, which is based on the density matching idea, offers a fast and stable algorithm that performs comparably to competing methods, while having a form that greatly simplifies the theoretical analysis of the data-driven procedure.

	Assume $supp(g_\mu)\subset [-M,M]$, $M<\infty$. The $g$-modeling approach
	\citep{jiang2009general,koenker2014convex} involves approximating $g_\mu(\cdot)$ using a mixture of point masses. We form a grid of size $k$ evenly spaced between $-M$ and $M$: $$\left\{s, s+\eta, s+2\eta,..., s+(k-1)\eta\right\},$$ where $\eta=2M/(k-1)$ and $s=-M$. 
	Then $g_\mu(\cdot)$ can be approximated by {$\hat{g}_\mu(\cdot)=\sum_{j=1}^{k}w_jI_{s+(j-1)\eta(\cdot)}$}, where $I_c(\cdot)$ is the Dirac delta function centered at $c$. The task at hand then boils down to determining the optimal weights {$\pmb{w} = (w_1, w_2, ..., w_k)$}, which can be efficiently solved through a direct optimization approach.

	We outline a ``density matching'' approach for formulating the optimization objective function. Specifically, two different techniques are employed to derive the density estimate, namely, $\hat f$, which is constructed based on a given $\hat g$, and $\hat f^m$, which is constructed using a weighted bivariate kernel estimator. The objective function is designed to ensure a high degree of similarity between the two density estimators.  
	
	First, upon obtaining $\hat g$, a natural estimate for $f_i(x_i)$ is readily provided by
	$$
	\hat{f}_i(x)=\int_{-\infty}^{\infty}\phi_{\sigma_i}(x-y)\hat{g}_{\mu}(y)dy=\sum_{j=1}^{k}w_j\phi_{\sigma_i}\left\{x-s-(j-1)\eta\right\}.
	$$ 
	On the other hand, $f_i$ can be estimated by employing a weighted bivariate kernel estimator:
	$$
	\hat{f}^m_i(x)=\sum_{j=1}^{m}\dfrac{\phi_{h_\sigma}(\sigma_i-\sigma_j)  }{\sum_{k=1}^{m}\phi_{h_\sigma}(\sigma_i-\sigma_k) }\phi_{h_{xj}}(x-x_j),
	$$
	where ${\bm h}=(h_x, h_{\sigma})$ is a pair of bandwidths,  $\phi_{h_{\sigma}}(\sigma-\sigma_j)/\{\sum_{j=1}^m\phi_{h_{\sigma}}(\sigma-\sigma_j)\}$ determines the contribution of  $(x_j, \sigma_j)$ based on $\sigma_j$, $h_{xj}=h_x\sigma_j$ is a bandwidth that varies across $j$, and $\phi_{h}(z)= (1/\sqrt{2\pi h^2})\exp\left\{-z^2/(2h^2)\right\}$ is a Gaussian kernel. The motivation behind this approach is to leverage the smooth variation of $f_i(x_i)$ with respect to $\sigma_i$. The variable bandwidth $h_{xj}$ is utilized to account for the heteroscedasticity inherent in the data, resulting in data points with higher variation being associated with flatter kernels.

	To minimize the discrepancy between $\hat f_i$ and $\hat f_i^m$, we aim to find $\pmb{w}$ that solves the following convex optimization problem:
	\begin{equation}\label{deconv:opt}
		\mbox{Minimize}\ \  \sum_{i=1}^{m}\{\hat{f}_i(x_i)-\hat{f}^m_i(x_i)\}^2 \quad
		\mbox{subject\ to} \quad w_j\geq 0 \mbox{ for } 1\leq j\leq k \mbox{ and } \sum_{j=1}^{k}w_j=1.
	\end{equation}

	Denote $\pmb w^*=(w_1^*,\ldots, w_k^*)$ the optimizer of \eqref{deconv:opt}, then $\hat{f}_{0i}$ and $\hat{f}_i$ can be computed as 
	\begin{eqnarray*}
		\hat{f}_{0i}(x) & = & \int_{-\infty}^{\mu_0}\phi_{\sigma_i}(x-\mu)\hat{g}_\mu(\mu)d\mu=\sum_{s+(j-1)\eta\leq \mu_0} w^*_j\phi_{\sigma_i}(x-s-j \eta) \mbox{ and } \\
		\hat{f}_{i}(x) & = & \int_{-\infty}^{\infty}\phi_{\sigma_i}(x-\mu)\hat{g}_\mu(\mu)d\mu=\sum_{j=1}^{k} w^*_j\phi_{\sigma_i}(x-s-(j-1)\eta).
	\end{eqnarray*}
	Finally, $S_i$ can be estimated correspondingly using a plug-in method. 
	
	\subsection{A computational shortcut and the step-wise algorithm}\label{shortcut}
	
	The oracle rule necessitates a search over a two-dimensional space, denoted by $\{(c_1, c_2): \left[c^-_1 ,c^+_1\right] \times \left[ c^-_2 , c^+_2\right] \}$, for identifying the optimal cutoffs $(c_1^{OR}, c_2^{OR})$ defined in \eqref{opt-cutoffs}. This task can be computationally demanding. To overcome this challenge, we propose in this section a computational shortcut that leads to a considerable improvement in computational efficiency.

	To maximize $\mbox{ETP}^*(c_1,c_2)$ for a given $c_2$, our strategy must reject as many hypotheses as possible from group 1. This involves the selection of the smallest $c_1$ that satisfies the mFDR constraint. Consequently, the optimal solution $(c^{OR}_1,c^{OR}_2)$  must be located on the one-dimensional curve $$L(c_2)=\{(c_1^*(c_2), c_2),  c^-_2\leq c_2\leq c^+_2\},$$ where $c_1^*(c_2)=\inf\{c_1 \in \left[c^-_1 ,c^+_1\right]: \mbox{mFDR}_{\pmb{\delta}(c_1,c_2)}\leq \alpha\}$. The problem boils down to determining the optimal $c_2$ on $L(c_2)$ such that $\text{ETP}^*(c_2; L)\equiv \text{ETP}^*(c_1^*(c_2), c_2)$ can be maximized. The following proposition establishes that if $\text{ETP}^*(c_2; L)$ starts to decrease along the curve $L(c_2)$ in the direction of increasing $c_2$, then it will continue to decrease in the direction of increasing $c_2$. 
	
	\begin{proposition}\label{monotone} Consider three decision rules $\pmb{\delta}=\pmb{\delta}(c_1,c_2)$, $\pmb{\delta}'=\pmb{\delta}(c'_1,c'_2)$, $\pmb{\delta}''=\pmb{\delta}(c''_1,c''_2)$ of the form described in \eqref{opt-rule} with $(c_1,c_2),\ (c'_1,c'_2),\ (c''_1,c''_2) $ all on $L$. If 
		$c_1\geq c'_1\geq c''_1$ and $\mbox{ETP}^*_{\pmb{\delta}}\geq \mbox{ETP}^*_{\pmb{\delta}'}$, then we must have $\mbox{ETP}^*_{\pmb{\delta}'}\geq \mbox{ETP}^*_{\pmb{\delta}''}$. 
	\end{proposition}

	Proposition \ref{monotone} inspires us to adopt the following strategy: searching along the curve $L$ in the direction of increasing $c_2$ and stopping when $\text{ETP}^*$ begins to decrease. More precisely, we first select as many units as possible from group 1 and record the resulting $\mbox{ETP}^*$ (Step 3). Next, we select a single hypothesis from group 2 (Step 4), which decreases the $\mbox{ETP}^*$ but increases the FDR capacity. We then return to Step 3 and select as many units as possible from group 1 using the additional FDR capacity, and record the new $\mbox{ETP}^*$. We compare the new $\mbox{ETP}^*$ with the previous $\mbox{ETP}^*$. If the $\mbox{ETP}^*$ increases after the iteration, we repeat the aforementioned process (e.g. continue to Step 4 and return to Step 3), otherwise we stop the procedure and output the thresholds. The operation of the step-wise data-driven procedure is detailed in Algorithm \ref{algorithm.table}.

	
	\begin{algorithm}[!ht]
		\DontPrintSemicolon
		\textbf{Input}: {observations $\{x_i, i\in[m]\}$, estimated Clfdr values $\widehat{\pmb{\mbox{Clfdr}}}=\{\hat{\mbox{Clfdr}}_i, i\in[m]\}$, pre-specified reference level $\mu_0$ and FDR level $\alpha$.}
		
		\textbf{Output}: {The indices of selected study units.}

		\textbf{Step 1: } Compute $\hat{T}_i=(x_i-\mu_{0})/(\widehat{\mbox{Clfdr}}_i-\alpha),$ set $ \hat{S}_i=\xi(\hat{T}_i)$. Form the 4 groups described in the oracle procedure using $\widehat{\pmb{\mbox{Clfdr}}}$ in place of $ \pmb{\mbox{Clfdr}}$.  
		\medskip
		
		\textbf{Step 2:} Let $\mathcal{R}$ be the present collection of selected units, which  encompasses all units in group 0. 
		\medskip
		
		\textbf{Step 3: } 
		For units in group 1, arrange them in descending order of the estimated $\hat{S}_i$ and
		denote the ranked hypotheses by $H^1_{(1)},H^1_{(2)},...$ with the corresponding Clfdr designated as $\mbox{Clfdr}^1_{(1)},\mbox{Clfdr}^1_{(2)},...$. Let $k=\max\{j: \sum_{i=1}^{j}(\mbox{Clfdr}^1_{(i)}-\alpha)\leq -\sum_{i\in\mathcal{R}}(\mbox{Clfdr}_i-\alpha)\}$, and place $H_{(1)},H_{(2)},... H_{(k)}$ in the present selection set $\mathcal{R}$. 
		Compute and store $\mbox{ETP}^*=   \sum_{i\in \mathcal{R}} (x_i-\mu_{0}).$
		\medskip
		
		\textbf{Step 4: } For units in group 2, arrange them in ascending order of $\hat{S}_i$. Denote the ranked hypotheses by $H^2_{(1)},H^2_{(2)},...$. Identify the smallest integer $j$ such that $H^2_{(j)}$ has not been incorporated into $\mathcal{R}$, and subsequently append $H^2_{(j)}$ to $\mathcal{R}$. 
		
		\medskip
		
		\textbf{Step 5: } Proceed to repeat step 3 and step 4. Terminate the process when $\mbox{ETP}^*$ begins to decline or when either all units in group 1 or group 2 have been selected.
		
		\medskip
		
		\textbf{Step 6: } Terminate the algorithm  if all units in group 1 or group 2 have been selected.  Return $\mathcal{R}$.
		
		\textbf{Step 7: } If $\mbox{ETP}^*$ starts to decrease, denote $H^2_{(j)}$ as the last unit added to $\mathcal{R}$ from group 2. Establish $\tilde{\mathcal{R}}$ by allocating all units in group 0 into $\tilde{\mathcal{R}}$, followed by $H^2_{(1)},\ldots, H^2_{(j-1)}$. Next, obtain $k=\max\{j: \sum_{i=1}^{j}(\mbox{Clfdr}^1_{(i)}-\alpha)\leq -\sum_{i\in\tilde{\mathcal{R}}}(\mbox{Clfdr}_i-\alpha)\}$, and insert $H_{(1)},H_{(2)},... H_{(k)}$ into $\tilde{\mathcal{R}}$. Return $\tilde{\mathcal{R}}$.

		\caption{The data-driven procedure}\label{algorithm.table}
	\end{algorithm}

	\subsection{Theoretical properties of the data-driven procedure}\label{theory}

	In Section \ref{oracle}, we have demonstrated that the oracle rule $\pmb{\delta}^{OR}$ is both valid and optimal in the sense that it satisfies the FDR constraint and has the largest ETP* among all valid FDR rules. In this subsection, we aim to establish that the data-driven procedure $\pmb{\delta}^{DD}$, defined in Algorithm \ref{algorithm.table}, asymptotically approaches the performance of the oracle rule $\pmb{\delta}^{OR}$, and therefore is asymptotically valid and optimal. Before we proceed with our theoretical analysis, we state the following regularity conditions.

	\begin{description}	
		\item\textbf{(A1)} $supp(g_\mu)\subset [-M,M]$ and $supp(g_\sigma)\in (M_1,M_2)$ for some $M_1>0$, $M_2<\infty$, $M<\infty$.
		\item\textbf{(A2)} The bandwidths  ${\bm h}=(h_x, h_{\sigma})$ satisfy $h_x\sim m^{-\eta_x}$, $h_\sigma\sim m^{-\eta_s}$ where $\eta_x$ and $\eta_s$ are small positive constants such that $0<\eta_s+\eta_x<1$.

		\item \textbf{(A3)} The grid size satisfies $k \sim m^{1/3}\log m$. 
	\end{description}
	\begin{remark}

		Assumption (A1) is a reasonable requirement for the boundedness of $g_\mu$ and $g_\sigma$ in most practical scenarios. Similarly, Assumption (A2) is fulfilled by commonly utilized bandwidth choices in \cite{wand1996kernel}. Proper selections of grid size by users can satisfy Assumption (A3), which can be relaxed to $k\rightarrow \infty$ as $m\rightarrow \infty$. While a larger $k$ does not compromise the quality of the deconvolution estimate in theory, it may increase computational times. In Section \ref{grid}, we show that the grid size $k$ does not need to be of order greater than $m^{1/3}\log m$. The performance of the estimator is contingent on the rate of convergence of $\mathbb{E}\|\hat{f}^m_\sigma-f_\sigma\|_2^2$. With appropriate selections of $h_x$ and $h_\sigma$, the fastest rate of convergence for $\mathbb{E}\|\hat{f}^m_\sigma-f_\sigma\|_2^2$ is $m^{-2/3}$. A sufficient grid size $k$ is one that allows $\hat{f}$ to approximate $\hat{f}^m$ at a rate of $m^{-2/3}$ in integrated mean squared error.

	\end{remark}

	We first state a crucial proposition that establishes the theoretical properties of the proposed density estimator $\hat{f}_{0i}$.
	\begin{proposition}\label{prop1}
		Suppose condition (A1), (A2), and (A3) hold, then
		$\hlfdr_i\overset{p}{\rightarrow}\mbox{Clfdr}_i$ when $m\rightarrow\infty$.
	\end{proposition}

	Now we present our theory on the asymptotic validity and optimality of the data-driven prioritized selection procedure. 
	\begin{theorem}\label{dd}
		Under Conditions (A1), (A2) and (A3), the data-driven procedure $\pmb{\delta}^{DD}$ described in Algorithm \ref{algorithm.table} controls mFDR at level $\alpha+o(1)$ and $\mbox{ETP}^*_{\pmb{\delta}^{DD}}/\mbox{ETP}^*_{\pmb{\delta}^{OR}}= 1+o(1)$ as $m\rightarrow \infty$.
	\end{theorem}

	\section{The R-Value in Multiple Comparisons}\label{sec:r-value}
	
	In this section, we investigate the integration of ranking and selection in a unified multiple comparison framework. Our proposed approach involves generating a ranking based on a suitable selection rule, and utilizing a novel ranking metric called the r-value. This metric reflects the relative order in which different units are selected, thereby providing a practical criterion for assessing the relative importance of the units within a list. 
	
	We present two r-value notions. The first, presented in Section \ref{subsec:R-value1}, discusses the situation where the reference level $\mu_0$ is fixed and the r-values are generated by varying the confidence level $\alpha$. The second, presented in Section \ref{subsec:R-value2}, discusses the situation where the confidence level $\alpha$ is fixed and the r-values are generated by varying the reference level $\mu_0$. Important properties of the r-values are investigated in Section \ref{subsec:agreeability}.  
	
	\subsection{R-values generated by varying the confidence level}\label{subsec:R-value1}
	
	The conventional multiple testing framework relies on a thresholding procedure that assumes the presence of a significance index, such as the $p$-value or local false discovery rate, which provides a consistent ranking of study units that remains invariant across all FDR levels. However, in the case of a heteroscedastic setup, such a ranking cannot be provided. For instance, in the oracle rule \eqref{opt-rule}, study units may be selected into the rejection set in different orders at varying FDR levels since the optimal statistic $T$ depends on $\alpha$. As a result, the ranking would be inconsistent across different users who may select different FDR levels in their analysis. Furthermore, there is no natural order for the subjects in group 0, as all units in the whole group are selected simultaneously. 
	
	Next we propose the pivotal notion of r-value, which has the ability to convert any selection procedure that controls the error probability into a meaningful and coherent ranking metric.
	\begin{definition}\label{rval}
		Let $\mathcal{R}^\mathcal{D}_\alpha$ denote the set of units selected by a pre-defined selection procedure $\mathcal{D}$ that controls the error rate at level $\alpha$. The r-value of a unit $i \in [m]$ linked with $\mathcal{D}$ is defined as 
		\begin{equation}\label{def:rv1}	
			r_i = \inf\{\alpha: i\in \mathcal{R}^\mathcal{D}_\alpha\}.
		\end{equation}
	\end{definition} 
	
	\begin{remark}
		In Supplementary Material \ref{sec:rv-pv-qv}, we demonstrate that the r-value defined by \eqref{def:rv1} encompasses the conventional p-value and q-value as special cases, provided that meaningful error concepts and their corresponding selection procedures are appropriately employed. 
	\end{remark}
	
	When combined with the novel prioritized selection procedure that solves \eqref{eq:RS-problem2}, the r-value corresponds to the minimum FDR level at which a study unit can be selected. This ranking metric addresses the inconsistency issue that may arise from the subjective specification of $\alpha$ values, offering an objective and consistent means of ranking across different users.
	
	\subsection{R-values generated by varying the reference level}\label{subsec:R-value2}
	
	The specification of $\mu_0$ in practical scenarios hinges on prior domain expertise, which may be subjective and vary among users. Since the oracle statistic $T$ \eqref{T-OR} and the corresponding data-driven quantity are contingent on $\mu_0$, divergent selections of $\mu_0$ among analysts may yield inconsistent rankings. Assuming a consensus on the choice of the confidence parameter $\alpha$ (e.g., 0.05), it is possible to generate r-values by varying the reference level $\mu_0$. Suppose we vary $\mu_0$ from $\infty$ to $-\infty$, then the earlier a unit is selected, the more important it is considered to be relative to the other units -- thus an objective ranking of study units is generated. 
	
	\begin{definition}\label{rval2}
		Let $\mathcal{R}^\mathcal{D}_{\mu_0}$ denote the set of units selected by a pre-defined selection procedure $\mathcal{D}$ that aims to select units with effect size larger than $\mu_0$. The r-value of a unit $i \in [m]$ associated with $\mathcal{D}$ is defined as follows:
		$$
		r_i= \sup\{\mu_0: i\in \mathcal{R}^\mathcal{D}_{\mu_0}\}, \text{ or } r_i^\prime=\frac{1+\sum_{j\neq i}\II(r_j>r_i)}{m},
		$$
		provided that no ties exist between $r_i$'s. 
	\end{definition}
	Here, $r_i^\prime\in \{i/m: i\in[m]\}$ is the standardized rank taking values in $(0,1]$, which is suitable in situations where only the relative position of the study units is relevant. The non-standardized r-value of a particular unit $i$ corresponds to the largest predetermined reference value $\mu_0$ at which unit $i$ can be selected with confidence.
	
	\subsection{Which r-value to use}
	
	By integrating our r-value with the prioritized selection framework \eqref{eq:RS-problem2}, we have developed a solution that is both intuitive and logically coherent for the challenging problem of ranking and selection under heteroscedastic setups. Both definitions of the r-value have their own advantages and drawbacks. To generate the r-value in this context, one can vary either $\alpha$ or $\mu_0$, depending on whether a consensus has been reached on the confidence level or reference level. 
	
	It is worth noting that Definition 2 of the r-value appears to align more closely with the primary objective of the two applications presented in Section \ref{data}, which is to identify schools/funds where there exists a substantial gap between two groups, with a predetermined level of confidence. By adopting Definition 2, we are able to circumvent the challenge of establishing a suitable gap threshold, which can be subjective in practice due to, say, the variability in passing rates between schools and the absence of a clear benchmark difference. Rather than focusing on the specific threshold value, our approach aims to detect substantial effect gaps with confidence, which better aligns with the practical needs of the analysis.
	
	\subsection{Agreeability of ranking}\label{subsec:agreeability}

	The proposed framework of ``select and then rank'' offers an appealing alternative to the conventional approach of ``rank and then select,'' which is impractical in the presence of heteroscedastic data. For example, Definition 1 first tackles the selection issue through constrained optimization, resulting in an objective solution for any given $\alpha$. Next, the ranking issue is handled using the r-value, which is determined by sequentially adjusting the selection level $\alpha$ without any user input. Consequently, the needs for a universally applicable test statistic and a potentially subjective choice of $\alpha$ can be eliminated, thus preventing the issue of inconsistent rankings.

	To demonstrate the appropriateness of the ranking generated by r-values, we introduce the concept of \emph{agreeability}. As previously mentioned, the ranking in heteroscedastic scenarios must consider two factors: effect size (captured by $X$) and statistical significance (captured by Clfdr or its estimate $\widehat{\mbox{Clfdr}}$). The following theorem asserts that if unit $i$ dominates unit $j$ in terms of both effect size and statistical significance, then the use of the r-value ensures that unit $i$ will be ranked higher than unit $j$.

	\begin{theorem}\label{consistent}
		Let $r_i$ and $\hat{r}_i$ be the r-values produced by the oracle procedure \eqref{opt-cutoffs} and the data-driven procedure (Algorithm 1), respectively, for $i\in[m]$. Then both the oracle and data-driven procedures are agreeable in the sense that if $X_i>X_j$ and $\operatorname{Clfdr}_i<\operatorname{Clfdr}_j$ (or $\widehat{\operatorname{Clfdr}}_i<\widehat{\operatorname{Clfdr}}_j$), then $r_i<r_j$ (or $\hat{r}_i<\hat{r}_j$). This assertion holds true for both Definition 1 and Definition 2 with $r^\prime_i$. 
	\end{theorem}
	
	\begin{remark}\rm{
			Agreeability  can be seen as a less stringent version of the nestedness notion. \cite{GuKoenker2020invidious}  explored the notion of nestedness  in ranking and selection, while \cite{HendersonNewton2014} suggested some potential issues regarding the nestedness requirement in the presence of heteroscedasticity. In Section \ref{subsec:nest} of the Supplementary Material, we precisely define the nestedness property and present counterexamples to demonstrate why nested selection may be infeasible under heteroscedastic setups. }
	\end{remark}

	\section{Numeric experiments}\label{sec:numerical}

	We begin by presenting the implementation details of the data-driven procedure in Section \ref{implementationAndSimu1}. To demonstrate the effectiveness of the modified power function, Section \ref{simu:3} provides a comparative analysis of the ETP* and ETP across several settings. In Section \ref{simu1}, we investigate the performance of the prioritized selection procedure and compare it with competing methods in a scenario where both $g_\sigma(\cdot)$ and $g_\mu(\cdot)$ are continuous. In Section \ref{simu:2}, we present additional results for the case where both $g_\sigma(\cdot)$ and $g_\mu(\cdot)$ are discrete, and where $\mu_i$ is correlated with $\sigma_i$.
	
	\subsection{Some implementation details}\label{implementationAndSimu1}

	The nonparametric deconvolution method discussed in Section \ref{deconv} requires the estimation of $\hat{f}_i^m(x_i)$, $i\in[m]$, which involves specifying the tuning parameters $\pmb{h}=(h_x,h_\sigma)$. In our analysis, we have employed the rule of thumb in \citet{Sil86}, which suggests $h_x=0.9\min\{\text{sd}(\pmb{x}), \text{IQR}(\pmb{x})  \}/(1.34m^{1/5})$ and $h_\sigma=0.9\min\{\text{sd}(\pmb{\sigma}), \text{IQR}(\pmb{\sigma})  \}/(1.34m^{1/5})$, where $\text{sd}(\cdot)$ and $\text{IQR}(\cdot)$ are the standard deviation and interquartile range of the input vector, respectively. In all our simulation studies, we have chosen a grid size $k$ of 50.
	
	To ensure numerical stability, we suggest selecting the support of the grid to be $[\hat{F}^{-1}(0.01),\hat{F}^{-1}(0.99)]$, where $\hat{F}^{-1}(\cdot)$ represents the empirical quantile function of $X_i$. We solve the convex optimization problem \eqref{deconv:opt} using the \texttt{CVXR} package in R \citep{cvxr}.
	
	\subsection{A comparative analysis of ETP* and ETP}\label{simu:3}
	
	In this section, we conduct numerical studies to demonstrate that maximizing ETP and ETP* are two distinct objectives. We generate $m=10000$ observations that follow the hierarchical model described below. 
	\begin{eqnarray*}
		\mu_i\sim N(5,0.5^2),  & \sigma_i=1,\quad X|\mu_i,\sigma_i\sim N(\mu_i,\sigma^2_i),\quad 1\leq i\leq 5000, \\
		\mu_i\sim N(7,0.5^2),  &\  \sigma_i=\sigma,\quad X|\mu_i,\sigma_i\sim N(\mu_i,\sigma^2_i),\quad 5001\leq i\leq 10000.
	\end{eqnarray*}
	
	We aim to test the hypotheses $H_{0,i}: \mu_i\leq \mu_0$ versus $H_{a,i}: \mu_i > \mu_0$, for $i\in[m]$, where $\mu_0=6$. We compare the performance of the following three methods:
	\begin{description}	
		\item (a) the oracle procedure derived in Section \ref{oracle}, denoted as OR; 
		\item (b) the data-driven method proposed in Section \ref{sec:dd}, denoted as DD; 
		\item (c) The oracle procedure designed to maximize the conventional ETP while controlling the FDR, denoted as Clfdr (see \cite{fu202hart} for further details). 
	\end{description}	
	
	We repeat the experiment on 100 datasets, and set the nominal FDR level to $\alpha=0.1$, and report the results based on the average of the 100 replications. The data-driven method requires the independence between $\sigma_i$ and $\mu_i$. To ensure the validity of the data-driven approach, we first partition the data into two groups based on whether $\sigma_i=1$ or $\sigma_i\neq1$. We then estimate $g_\mu(\cdot)$ separately for each of the two groups.
	
\begin{figure}
		\includegraphics[width=6in]{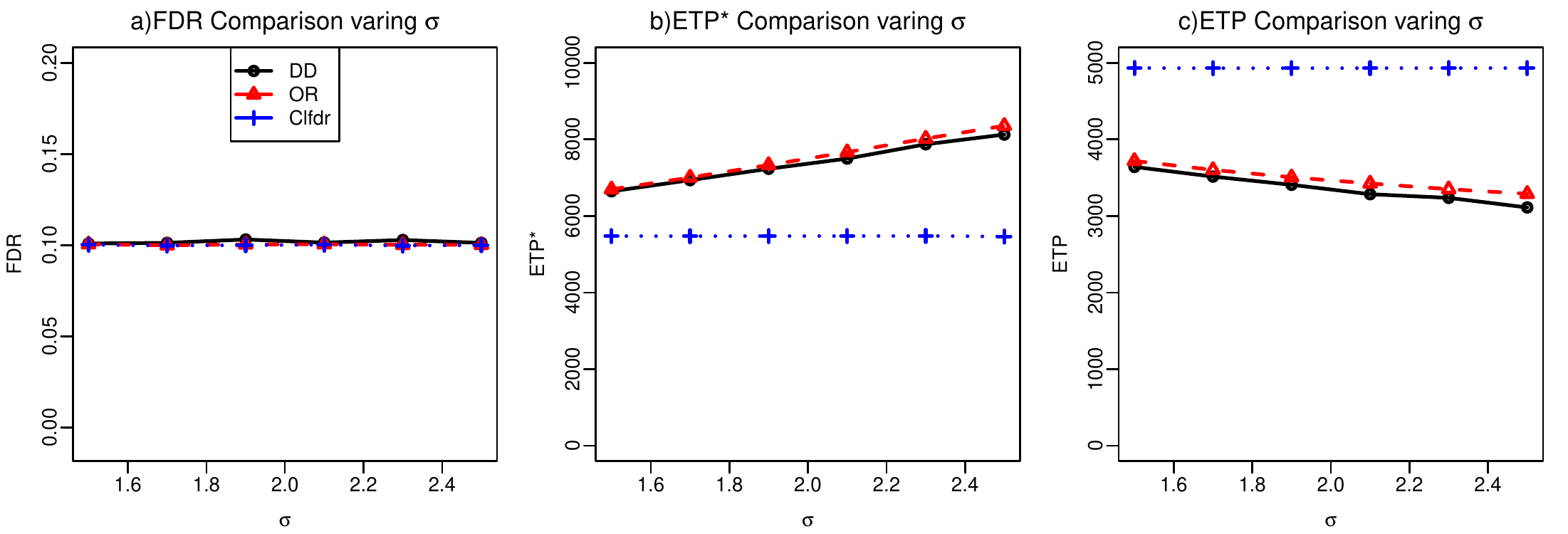} 
		\caption{An example where ETP* and ETP are very different. DD and OR have much higher ETP* than Clfdr while Clfdr has much higher ETP than DD and OR.  }
		\label{dd3}
	\end{figure}

	We calculate the FDR as the average of the FDPs over 100 replications. The FDP is defined as $\text{FDP}(\pmb{\delta})=\sum_{i=1}^{m}\{ (1-\theta_i)\delta_i \}/(\sum_{i=1}^{m}\delta_i\vee1)$. Similarly, we compute the ETP and ETP* as the averages of $\sum_{i=1}^{m}\theta_i\delta_i$ and $\sum_{i=1}^{m}(x_i-\mu_0)\delta_i$, respectively, over 100 replications. The value of $\sigma$ varies from 1.5 to 2.5 across different settings. We present the results in Fig \ref{dd3}.

	The results indicate that all three methods effectively control the FDR at the nominal level. However, there are significant differences in their power performance. In particular, the ETP* values of the OR and DD methods are substantially higher than that of Clfdr, while Clfdr exhibits a significantly higher ETP than OR and DD. This observation aligns with the fact that the Clfdr method is designed to optimize traditional power, whereas OR and DD are developed with the objective of optimizing modified power.
	
	\subsection{Comparison for independent $\mu_i$ and $\sigma_i$}\label{simu1}
	Next, we consider the following setting:
	$$
	\theta_i\overset{iid}{\sim}\mbox{Ber}(0.2),\quad \mu_i|\theta_i\sim (1-\theta_i)U(-3,-1)+\theta_iU(1,2),\quad \sigma_i\overset{iid}{\sim} U(0.5,\sigma_{max}),
	$$
	$$
	X_i|\mu_i,\sigma_i\sim N(\mu_i,\sigma^2_i),\quad i\in [5000].
	$$
	
\begin{figure}[t!]
		\includegraphics[width=6in]{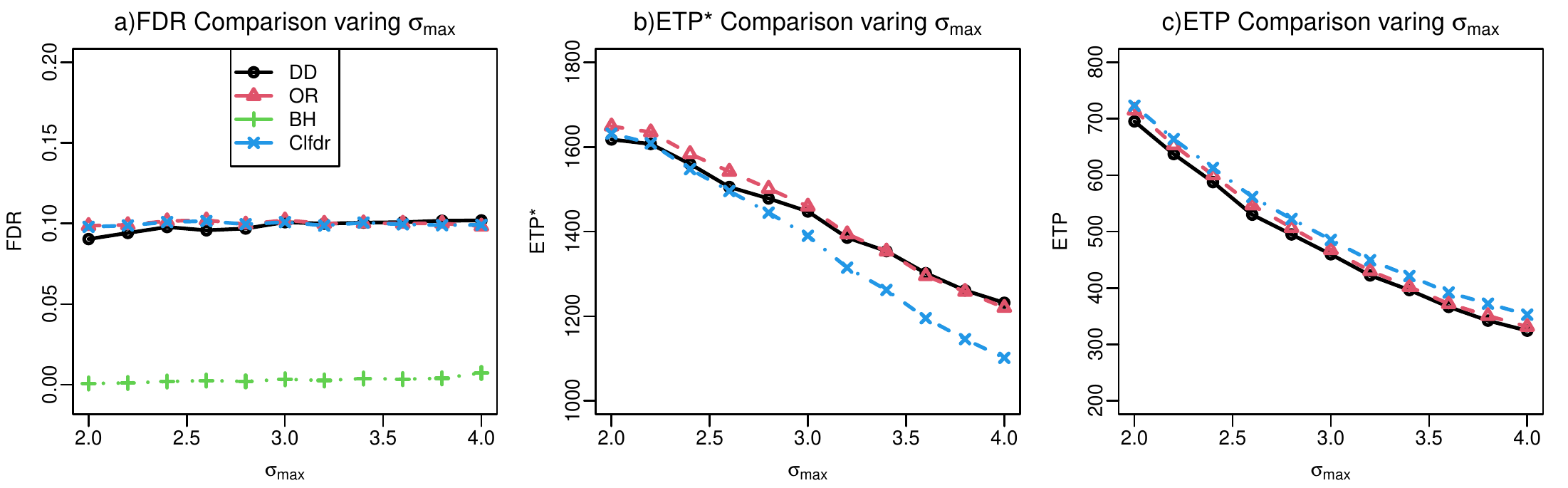} 
		\caption{\label{dx_sum}Comparison when $\sigma_i$ and $\mu_i$ are uncorrelated and both are generated from a uniform distribution. We vary $\sigma_{max}$ from 2 to 4. All methods control the FDR at the nominal level with BH being overly conservative. }
			\vspace{.5cm}
			\includegraphics[width=6in]{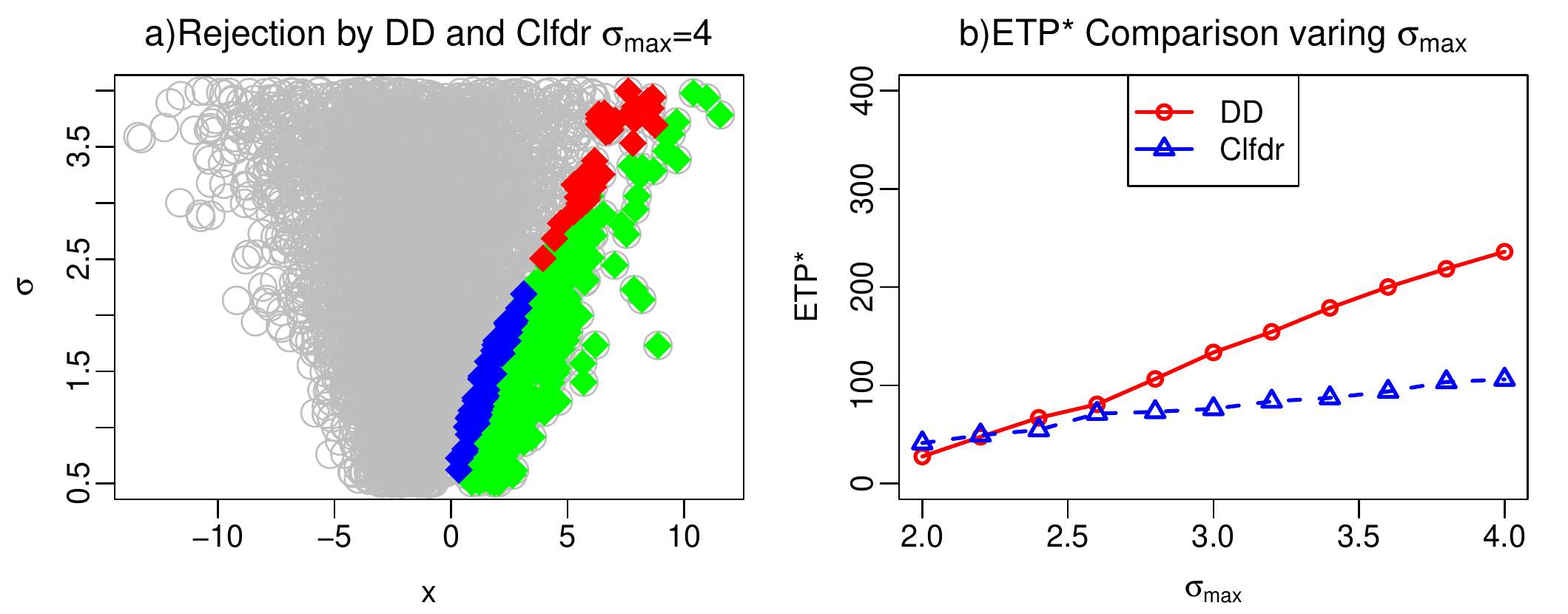} 
			\caption{(a): A scatter plot of the hypotheses when $\sigma_{max}=4$. The gray circles are hypotheses rejected by neither DD or Clfdr, green dots are hypotheses rejected by both DD and Clfdr, red dots are hypotheses rejected by DD but not Clfdr, blue dots are hypotheses rejected by Clfdr but not DD. (b): ETP* comparison of hypotheses rejected by either DD or Clfdr, but not both. }
			\label{dx_unifdiff}
	\end{figure}

	We aim to test the hypotheses $H_{0,i}: \mu_i\leq \mu_0$ versus $H_{a,i}: \mu_i > \mu_0$, with $\mu_0=0$. In addition to the three methods compared in Section \ref{simu:3}, we also incorporate the widely used Benjamini-Hochberg procedure (BH) procedure in the comparison. The p-values are computed as $1-\Phi\{ (X_i-\mu_0)/\sigma_i\}$, where $\Phi(\cdot)$ is the cumulative distribution function of a standard normal variable. The nominal FDR level is set to $\alpha=0.1$, while $\sigma_{max}$ varies from 2 to 4 for different settings. The results are obtained by averaging the results in 100 replications and are presented in Fig \ref{dx_sum}.

	We can observe two important patterns. Firstly, BH appears to be excessively conservative, suggesting that $p$-value based methods may not be well-suited for testing composite hypotheses. Secondly, DD, OR, and Clfdr exhibit comparable levels of FDR but display noticeable differences in their ETP* and ETP values.

	A more detailed comparison of the hypotheses rejected by DD and Clfdr underscores the marked differences between these two methods. In Fig \ref{dx_unifdiff} (a), we look at one particular run with  $\sigma_{max}=4$. The gray dots are hypotheses not rejected by either DD or Clfdr, green dots are hypotheses rejected by both DD and Clfdr, red dots are hypotheses rejected by DD but not Clfdr, and blue dots are hypotheses rejected by Clfdr but not DD. 
	
	Upon close examination, it is evident that DD is more likely to reject hypotheses with higher $x_i$ values when compared to Clfdr. If we exclude the hypotheses that are rejected by both DD and Clfdr and assess the ETP* for the remaining hypotheses, a distinct contrast emerges, as depicted in Figure \ref{dx_unifdiff} (b). For the hypotheses that are rejected by only one method, DD has a superior ETP* in comparison to Clfdr. Additionally, the difference in ETP* becomes more pronounced as the degree of heteroscedasticity increases.

	\subsection{Comparison for correlated $\mu_i$ and $\sigma_i$}\label{simu:2}
	
	In this section, we present simulation results in a more complex scenario where $\sigma_i$ and $\mu_i$ are correlated. Let $I_c$ be an indicator function that takes the value of $1$ at $c$ and $0$ elsewhere. The model first generates $\sigma_i$ from two groups and then generates $\mu_i$ in a manner that is dependent on $\sigma_i$, as described below: 
	\begin{eqnarray*}
		X_i|\mu_i,\sigma_i & \sim &  N(\mu_i,\sigma^2_i), \quad \sigma_i\overset{iid}{\sim}\frac{1}{2}I_{0.25\sigma}(\cdot)+\frac{1}{2}I_{1.25\sigma}(\cdot),  \\ 
		\mu_i|\sigma_i=0.25\sigma & \sim &  0.9N(-0.5,0.25^2)+0.1N(1.5,0.25^2), \\
		\mu_i|\sigma_i=1.25\sigma & \sim & 0.9N(-0.5,0.25^2)+0.1N(3,0.25^2).
	\end{eqnarray*}
	
	The hypotheses to be tested in our study are $H_{0,i}:\mu_i\leq \mu_0$ vs $H_{a,i}:\mu_i> \mu_0$, where $\mu_0=1, i\in[10000]$. 
	The value of $\sigma$ varies between 1.5 to 2 for different settings. 
	It is important to note that in our design, $\mu_i$ is sampled from a mixture distribution, where a vast majority of $\mu_i$ values are generated from $N(-0.5,0.25^2)$, corresponding to small effects. However, there is a small fraction of $\mu_i$ values that correspond to large effects. Furthermore, the effect sizes $\mu_i$ tend to increase as the variance becomes larger.

\begin{figure}
		\includegraphics[width=6in]{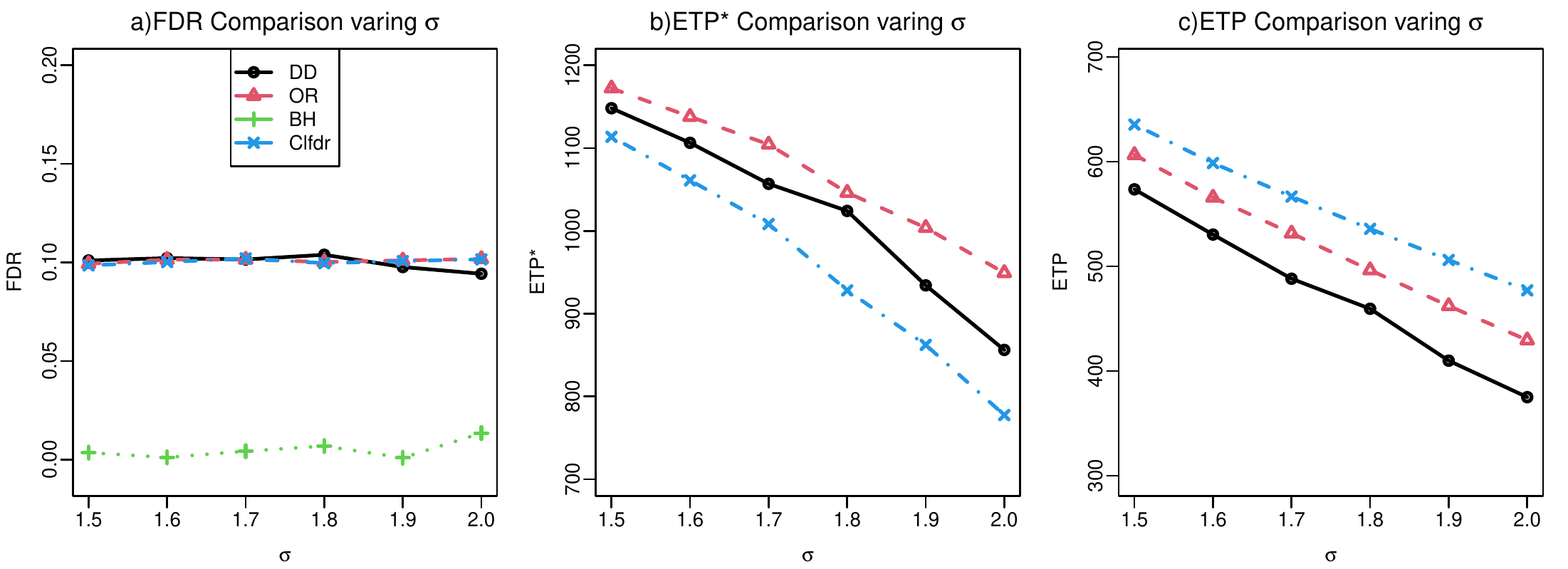} 
		\caption{The comparison under the setting where $\sigma_i$ and $\mu_i$ are correlated. Under the $\mbox{ETP}^*$ metric, DD and OR outperform Clfdr. Conversely, under the ETP metric, Clfdr exhibits a higher power than DD and OR. }
		\label{d_sum}
	\end{figure}

	We perform simulation experiments on 100 datasets and apply DD, OR, Clfdr, and BH to select important units at FDR level $\alpha=0.1$.
	The accuracy of the deconvolution estimation method relies on the independence between $\sigma_i$ and $\mu_i$. Therefore, we initially partitioned the data into two groups based on whether $\sigma_i=0.25\sigma$ or $\sigma_i=1.25\sigma$, and estimated $g_\mu(\cdot)$ separately for each group. A summary of results for different values of $\mu$ is presented in Figure \ref{d_sum}. 
	
\begin{figure}
		\includegraphics[width=6in]{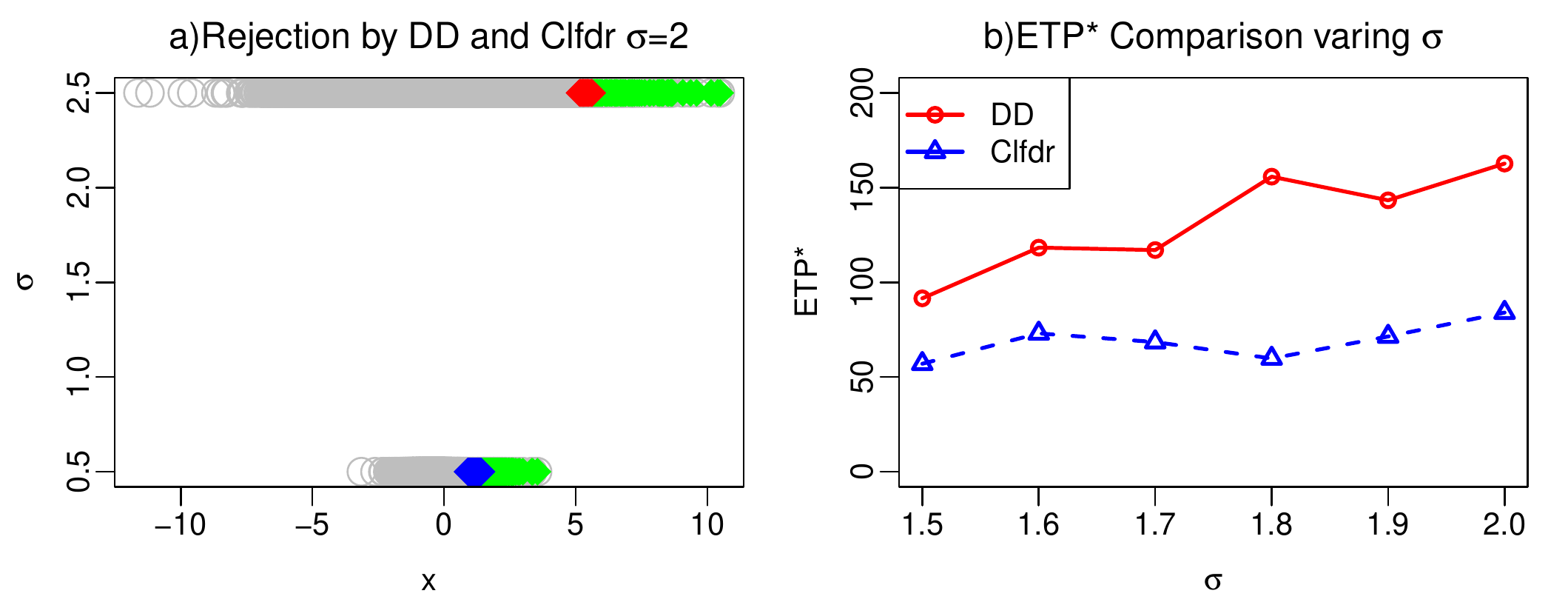} 
		\caption{(a): A scatter plot of the hypotheses when $\sigma=2$. The x-axis represents $x_i$ and the y-axis represents $\sigma_i$. The gray circles are hypotheses rejected by neither DD or Clfdr, green dots are hypotheses rejected by both DD and Clfdr, red dots are hypotheses rejected by DD but not Clfdr, blue dots are hypotheses rejected by Clfdr but not DD. (b): ETP* comparison on hypotheses not rejected by both DD and Clfdr. }
		\label{dsun_hist}
	\end{figure}

	Our analysis reveals several patterns. Firstly, all methods maintain FDR control at the nominal level. Secondly, BH is excessively conservative, resulting in ETP* and ETP values that are significantly lower than the other three methods. Therefore, we exclude BH from the ETP* and ETP plots. Thirdly, DD and OR outperform Clfdr in terms of the ETP* criterion, whereas Clfdr outperforms DD and OR regarding the ETP criterion. To make a further comparison between Clfdr and DD, we present a scatter plot of rejected hypotheses when $\sigma=2$ in Figure \ref{dsun_hist}. In Figure \ref{dsun_hist} (a), the hypotheses rejected by DD but not Clfdr all have $\sigma_i=2$. This implies that DD is more sensitive to larger variances than Clfdr. In Figure \ref{dsun_hist} (b), we observe that DD has a significantly higher ETP* on hypotheses where Clfdr and DD disagree.
	
	\section{Real Data Applications}\label{data}
	
	In this section, we analyze the test performance data of K-12 schools from the 2005 Annual Yearly Performance (AYP) study (Section \ref{sec:ayp}) and mutual fund data obtained from the Center for Research in Security Prices (CRSP) (Section \ref{subsec:CRSP}).
	
	\subsection{AYP data}\label{sec:ayp}
	
\begin{figure}[ht!]
		\includegraphics[width=6in]{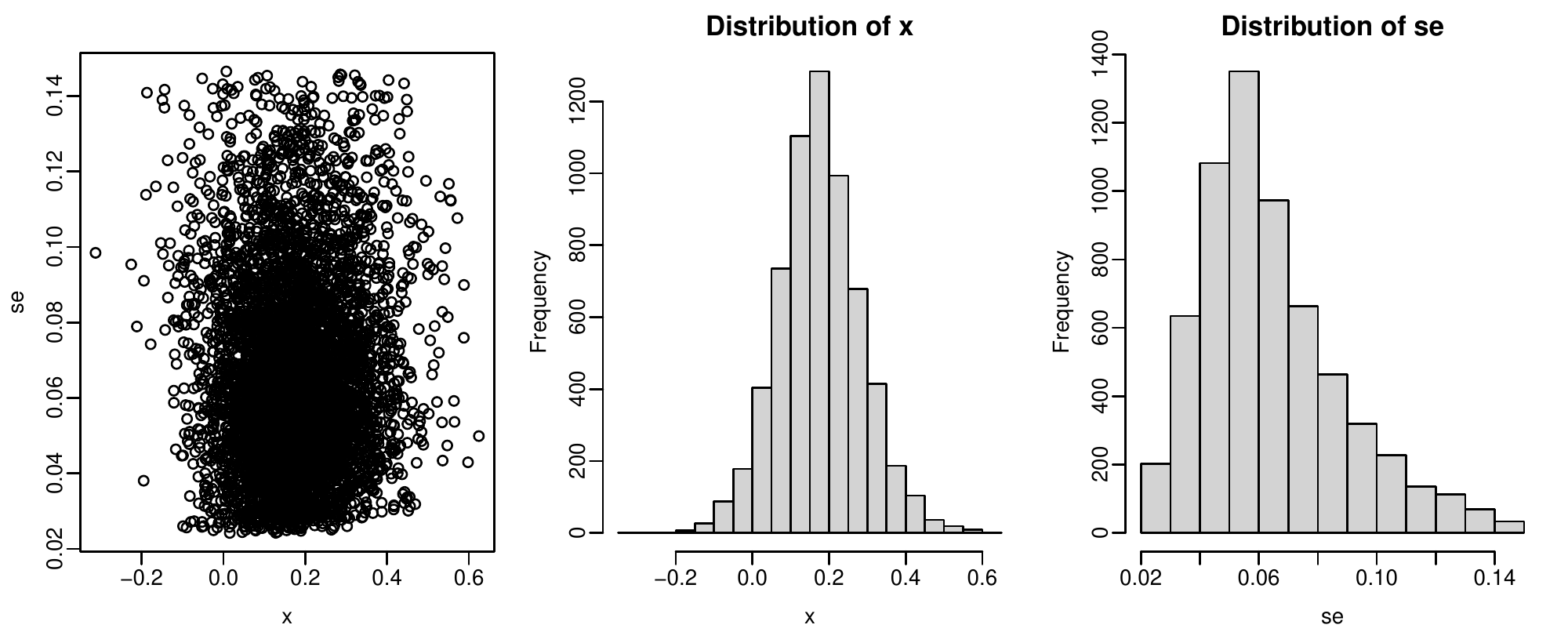}
		\caption{Left: scatter plot of the AYP data: x-axis is the observation, y-axis is the standard error. Middle: histogram of the observations. Right: histogram of the standard errors. }
		\label{sum}
	\end{figure}

	The unprocessed data sets for the AYP study are available at \url{https://www.cde.ca.gov/re/pr/api-datarecordlayouts.asp}. We begin by defining $Y$ and $Y'$ as the passing rates of students from socially-economically advantaged backgrounds (SEA) and socially-economically disadvantaged backgrounds (SED), respectively 
	Our objective is to identify significant differences in the passing rates $X_i = Y_i - Y'_i$ for each school $i$, where $i\in[m]$ and $m=6,398$. The standard error of $X_i$ is calculated as 
	$$s_i=\sqrt{ Y_i(1-Y_i)/{n_i}+{Y'(1-Y'_i)}/{n_{i'}} },$$
	where $n_i$ and $n_{i'}$ are the number of SEA and SED students tested, respectively. To ensure numerical stability, we remove observations that have a standard error below the $1\%$ percentile or above the $99\%$ percentile. Figure \ref{sum} presents the scatter plot and histograms of the observed data.
	
	We aim to test the following hypotheses: $H_{0,i}:\mu_i\leq \mu_0$ vs $H_{a,i}:\mu_i> \mu_0$, with $\mu_0=0.2$ being the cutoff of the indifference region and FDR level set at $\alpha = 0.01$. We calculate the z-values as $z_i=(x_i-\mu_0)/{s_i}$, and the p-values as $p_i=1-\Phi(z_i)$, where $\Phi$ is the standard normal cumulative distribution function. 
	
	Our primary focus is to compare our approach against analyses that solely rely on statistical significance indices (Clfdr and BH). In this context, the Clfdr method refers to the data-driven HART procedure \citep{fu202hart}. We summarize the results in Table \ref{table:realdata}, which reports the total number of rejections (a proxy for traditional power) and weighted number of rejections based on Equation \eqref{eq:RS-problem2} (a proxy for modified power). 
	
\begin{figure}[ht!]
		\includegraphics[width=6.5in]{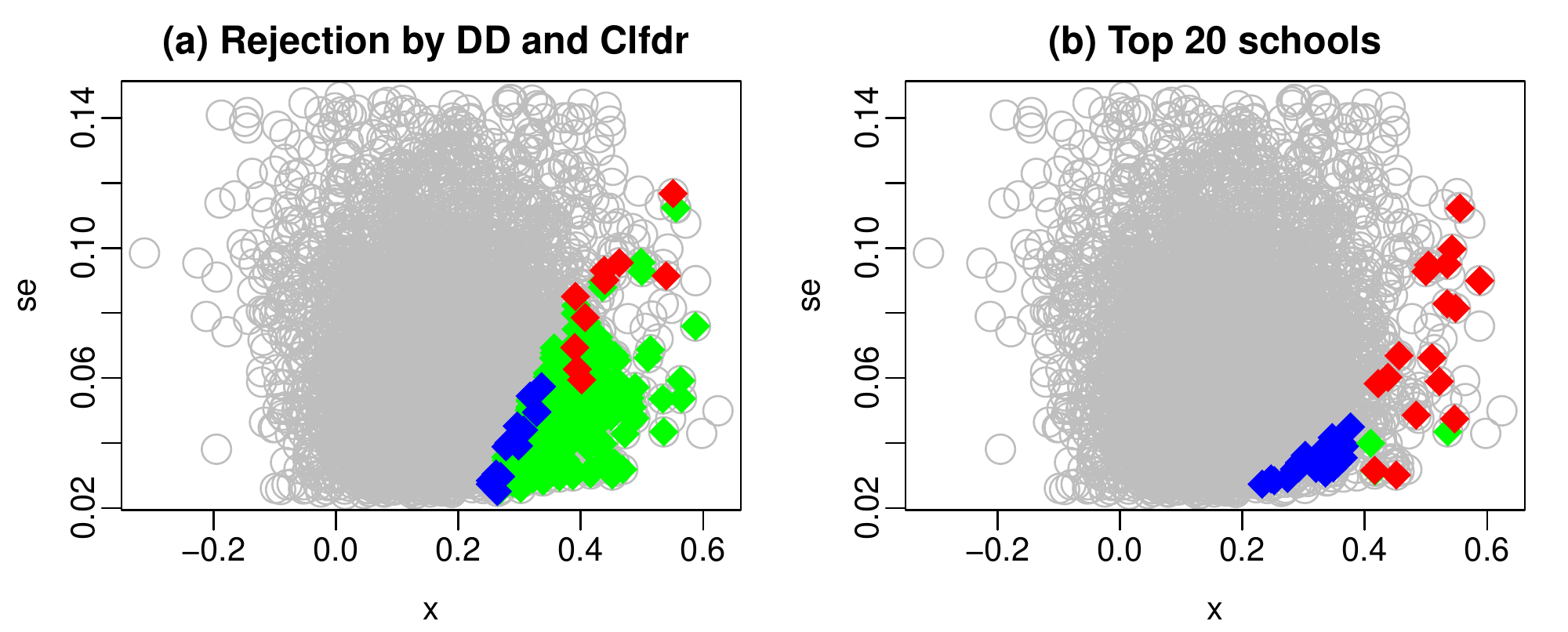} 
		\caption{(a) The scatter plot displays the school data. The x-axis represents the raw observations ($x$), while the y-axis corresponds to the SEs. The gray circles represent schools that were not selected by either DD or Clfdr. The green dots denote schools selected by both Clfdr and DD, while the blue dots represent schools selected by Clfdr but not DD. The hypotheses selected by DD but not Clfdr are shown as red dots. (b) The scatter plot displays the top 20 schools ranked according to p-value and r-value (Definition 2 with $\alpha=0.1$). The top 20 schools ranked according to the r-value are depicted as red dots, while the top 20 schools ranked according to the p-value are shown as blue dots. The schools ranked as top 20 by both p-value and r-value are represented by green dots. }
		\label{AYP_plot}
	\end{figure}

	We can see that DD and Clfdr outperform BH in terms of both the traditional and modified powers. Although DD and Clfdr exhibit similar performances, the hypotheses rejected by the two methods exhibit different patterns. Figure \ref{AYP_plot} (a) displays a scatter plot of hypotheses rejected by Clfdr and DD. The gray circles represent hypotheses that were not rejected by either method, the green dots represent hypotheses rejected by both Clfdr and DD, the red dots represent hypotheses rejected by DD but not Clfdr, and the blue dots represent hypotheses rejected by Clfdr but not DD. It is clear that DD displays a predilection for rejecting hypotheses with larger effect sizes, while Clfdr has a preference for rejecting hypotheses with low standard error. Figure \ref{AYP_plot} (b) presents the top 20 schools ranked according to both p-values (indicated by blue dots) and r-values (represented by red dots). Notably, the r-value demonstrates a distinct inclination towards schools with larger effect sizes as compared to the p-value.
	
	

	\begin{table}
		\caption{\label{table:realdata} Summary of power by each method on the AYP data.}
		\centering
		\begin{tabular}{|l|l|l|l|}
			\hline
			$ $& $\text{DD}$  & $ \text{Clfdr}$   & $ \text{BH}$\\
			\hline
			$\text{Number of hypotheses rejected}$ & $338$          &  $343$     &   $ 158$   \\
			\hline
			$\text{Modified Power}$ & $57.5$          &  $56.4 $     &   $ 34.3$   \\
			\hline
		\end{tabular}
		
	\end{table}
	
	\subsection{CRSP data}\label{subsec:CRSP}
	
	In this section, we analyze mutual fund data obtained from CRSP accessed via the Wharton WRDS database at the University of Pennsylvania.	The goal is to compare our ranking and selection method against Clfdr and BH that are solely based on significance indices.

\begin{figure}
		\includegraphics[width=6in]{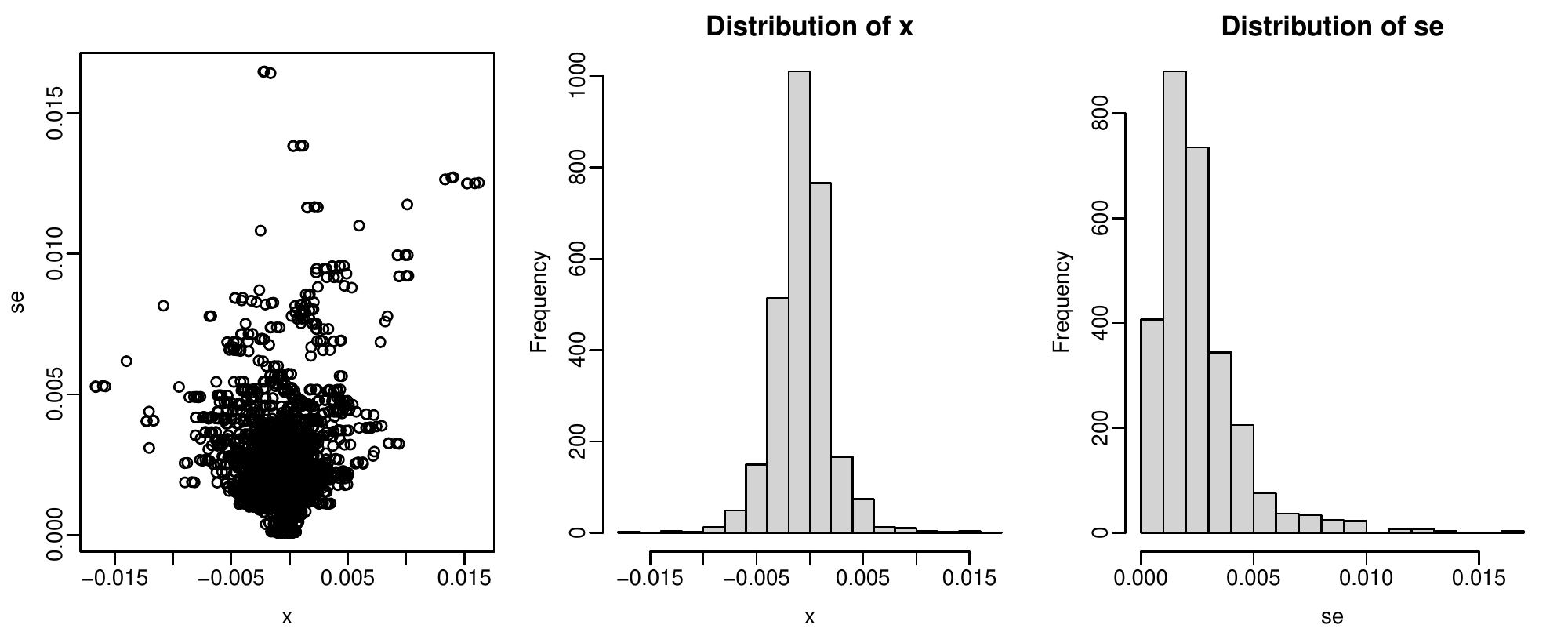} 
		\caption{Left: scatter plot of the CRSP data: x-axis is the observation, y-axis is the standard error. Middle: histogram of the observations. Right: histogram of the standard errors. }
		\label{sum2}
	\end{figure}

	We analyze the estimated returns of mutual funds, which are denoted as $x_i$, over an average of 31 months of performance from the end of April 2006 to the end of October 2008. These estimated returns are obtained from the intercept term of Carhart's four-factor model \citep{Carhart1997}. The standard error of the returns, denoted as $s_i$, is computed as the estimated standard error of the intercept term in the model. The dataset comprises $2796$ pairs of observations ($x_i$, $s_i$). To ensure numerical stability, we exclude observations with standard errors below the 0.1\% and above the 99.9\% percentiles. Figure \ref{sum2} visually depicts the distribution of the data.
	
	Our objective is to identify mutual funds that exhibit positive returns. Therefore, we consider the following hypotheses: $H_{0,i}:\mu_i\leq \mu_0$ vs $H_{a,i}:\mu_i > \mu_0$, with $\mu_{0}=0$. We set the target FDR level at $0.1$. The z-values and corresponding p-values are obtained as before.
	
	The results, presented in Table \ref{table:crsp}, reveal that Clfdr rejects more hypotheses than DD. However, Clfdr exhibits negative modified power, indicating its tendency to select portfolios with negative returns. This phenomenon occurs because Clfdr does not consider the value of the returns, leading it to select hypotheses with estimated returns ($x_i$) slightly lower than the null hypothesis ($\mu_0$), which corresponds to negative true returns. Clfdr selects such units because they do not increase the overall FDR above the target level. In practice, this type of ``over-selection'' is often undesirable, as demonstrated in this example.

	\begin{table}[ht!]
		\caption{\label{table:crsp} Summary of power by each method on the CRSP data.}
		\centering
		\begin{tabular}{|l|l|l|l|}
			\hline
			$ $& $\text{DD}$  & $ \text{Clfdr}$   & $ \text{BH}$\\
			\hline
			$\text{Number of hypotheses rejected}$ & $359$          &  $749$     &   $ 46$   \\
			\hline
			$\text{Modified Power}$ & $0.549$          &  $-0.282 $     &   $ 0.024$   \\
			\hline
		\end{tabular}
		
	\end{table}

\begin{figure}
		\includegraphics[width=6.5in]{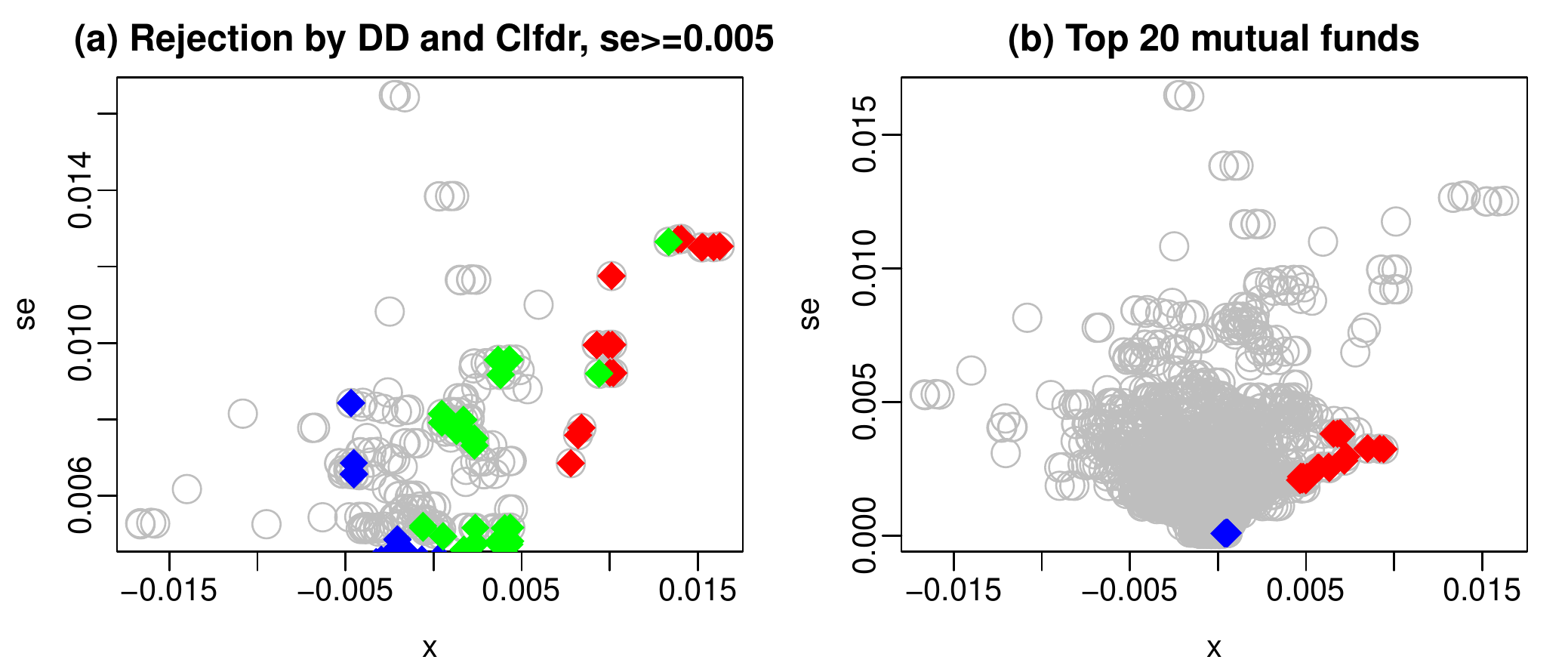} 
		\caption{(a) The scatter plot displays funds with an SE greater than 0.005. The x-axis represents observations ($x$), while the y-axis corresponds to the SEs. The gray circles denote funds that were not selected by either DD or Clfdr. The green dots represent funds selected by both Clfdr and DD, while the blue dots are funds selected by Clfdr but not DD and the red dots are funds selected by DD but not Clfdr. (b) The scatter plot displays the top 20 mutual funds ranked according to the p-value and r-value (Definition 2 with $\alpha=0.1$). The top 20 mutual funds ranked according to the r-value are denoted by red dots, while the top 20 mutual funds ranked according to the p-value are shown as blue dots.}
		\label{finance_plot}
	\end{figure}

	We proceed by examining the units selected by DD but not Clfdr, and vice versa. To provide a more informative comparison, we focus on hypotheses with $s_i\geq 0.005$. The results are presented in Figure \ref{finance_plot} (a). We observe that the units selected by DD and Clfdr with $s_i\geq 0.005$
	differ significantly. DD selects units with both high estimated returns ($x_i$) and high SEs, indicating its tendency to trade high variability for potentially high returns. Figure \ref{finance_plot} (b) displays the top 20 mutual funds ranked according to p-values (blue dots) and r-values(red dots). The results indicate that the r-value places higher priority on selecting funds with higher returns, whereas the p-value favors the selection of funds with smaller SEs.
	
\newpage

\bibliographystyle{chicago}
\bibliography{rsref.bib}

\begin{thebibliography}{}

\bibitem[\protect\citeauthoryear{Banerjee, Fu, James, and Sun}{Banerjee
  et~al.}{2020}]{Banetal21-NEST}
Banerjee, T., L.~J. Fu, G.~M. James, and W.~Sun (2020).
\newblock Nonparametric empirical bayes estimation on heterogeneous data.
\newblock {\em arXiv preprint arXiv:2002.12586\/}.

\bibitem[\protect\citeauthoryear{Basu, Cai, Das, and Sun}{Basu
  et~al.}{2018}]{Basuetal2018}
Basu, P., T.~T. Cai, K.~Das, and W.~Sun (2018).
\newblock Weighted false discovery rate control in large-scale multiple
  testing.
\newblock {\em Journal of the American Statistical Association\/}~{\em
  113\/}(523), 1172--1183.
\newblock PMID: 31011234.

\bibitem[\protect\citeauthoryear{Bechhofer}{Bechhofer}{1954}]{1954Bechofer}
Bechhofer, R.~E. (1954).
\newblock {A Single-Sample Multiple Decision Procedure for Ranking Means of
  Normal Populations with known Variances}.
\newblock {\em The Annals of Mathematical Statistics\/}~{\em 25\/}(1), 16 --
  39.

\bibitem[\protect\citeauthoryear{Benjamini and Hochberg}{Benjamini and
  Hochberg}{1995}]{bh1995}
Benjamini, Y. and Y.~Hochberg (1995).
\newblock Controlling the false discovery rate: a practical and powerful
  approach to multiple testing.
\newblock {\em Journal of the royal statistical society series
  b-methodological\/}~{\em 57\/}(1), 289--300.

\bibitem[\protect\citeauthoryear{Benjamini and Hochberg}{Benjamini and
  Hochberg}{2000}]{benjamini2000adaptive}
Benjamini, Y. and Y.~Hochberg (2000).
\newblock On the adaptive control of the false discovery rate in multiple
  testing with independent statistics.
\newblock {\em Journal of educational and Behavioral Statistics\/}~{\em
  25\/}(1), 60--83.

\bibitem[\protect\citeauthoryear{Boyd, Cortes, Mohri, and Radovanovic}{Boyd
  et~al.}{2012}]{2012Boydetal}
Boyd, S., C.~Cortes, M.~Mohri, and A.~Radovanovic (2012).
\newblock Accuracy at the top.
\newblock In F.~Pereira, C.~Burges, L.~Bottou, and K.~Weinberger (Eds.), {\em
  Advances in Neural Information Processing Systems}, Volume~25. Curran
  Associates, Inc.

\bibitem[\protect\citeauthoryear{Cai and Sun}{Cai and Sun}{2009}]{sun2009large}
Cai, T.~T. and W.~Sun (2009).
\newblock Simultaneous testing of grouped hypotheses: Finding needles in
  multiple haystacks.
\newblock {\em Journal of the American Statistical Association\/}~{\em
  104\/}(488), 1467--1481.

\bibitem[\protect\citeauthoryear{Cai, Sun, and Wang}{Cai
  et~al.}{2019}]{CaiSunWang2019}
Cai, T.~T., W.~Sun, and W.~Wang (2019).
\newblock Covariate-assisted ranking and screening for large-scale two-sample
  inference.
\newblock {\em Journal of The Royal Statistical Society Series B-statistical
  Methodology\/}~{\em 81}, 187--234.

\bibitem[\protect\citeauthoryear{Carhart}{Carhart}{1997}]{Carhart1997}
Carhart, M.~M. (1997).
\newblock On persistence in mutual fund performance.
\newblock {\em The Journal of Finance\/}~{\em 52\/}(1), 57--82.

\bibitem[\protect\citeauthoryear{Chen, Lin, Y\"{u}cesan, and Chick}{Chen
  et~al.}{2000}]{Chenetal2000}
Chen, C.-H., J.~Lin, E.~Y\"{u}cesan, and S.~E. Chick (2000, jul).
\newblock Simulation budget allocation for further enhancing the efficiency of
  ordinal optimization.
\newblock {\em Discrete Event Dynamic Systems\/}~{\em 10\/}(3), 251–270.

\bibitem[\protect\citeauthoryear{Efron}{Efron}{2011}]{Efron:2011:Tweedie}
Efron, B. (2011).
\newblock Tweedie’s formula and selection bias.
\newblock {\em Journal of the American Statistical Association\/}~{\em
  106\/}(496), 1602--1614.

\bibitem[\protect\citeauthoryear{Efron}{Efron}{2012}]{efron2012large}
Efron, B. (2012).
\newblock {\em Large-scale inference: empirical Bayes methods for estimation,
  testing, and prediction}, Volume~1.
\newblock Cambridge University Press.

\bibitem[\protect\citeauthoryear{Efron}{Efron}{2016}]{efron2016deconv}
Efron, B. (2016).
\newblock Empirical bayes deconvolution estimates.
\newblock {\em Biometrika\/}~{\em 103\/}(1), 1--20.

\bibitem[\protect\citeauthoryear{Efron, Tibshirani, Storey, and Tusher}{Efron
  et~al.}{2001}]{efron2001empirical}
Efron, B., R.~Tibshirani, J.~D. Storey, and V.~Tusher (2001).
\newblock Empirical bayes analysis of a microarray experiment.
\newblock {\em Journal of the American statistical association\/}~{\em
  96\/}(456), 1151--1160.

\bibitem[\protect\citeauthoryear{Foster and Stine}{Foster and
  Stine}{2008}]{foster2008alpha}
Foster, D.~P. and R.~A. Stine (2008).
\newblock $\alpha$-investing: a procedure for sequential control of expected
  false discoveries.
\newblock {\em Journal of the Royal Statistical Society: Series B (Statistical
  Methodology)\/}~{\em 70\/}(2), 429--444.

\bibitem[\protect\citeauthoryear{Fu, Gang, James, and Sun}{Fu
  et~al.}{2022}]{fu202hart}
Fu, L., B.~Gang, G.~M. James, and W.~Sun (2022).
\newblock Heteroscedasticity-adjusted ranking and thresholding for large-scale
  multiple testing.
\newblock {\em Journal of the American Statistical Association\/}~{\em
  117\/}(538), 1028--1040.

\bibitem[\protect\citeauthoryear{Gang, Sun, and Wang}{Gang
  et~al.}{2023}]{gang2021structure}
Gang, B., W.~Sun, and W.~Wang (2023).
\newblock Structure--adaptive sequential testing for online false discovery
  rate control.
\newblock {\em Journal of the American Statistical Association\/}~{\em
  118\/}(541), 732--745.

\bibitem[\protect\citeauthoryear{Genovese and Wasserman}{Genovese and
  Wasserman}{2002}]{genovese2002operating}
Genovese, C. and L.~Wasserman (2002).
\newblock Operating characteristics and extensions of the false discovery rate
  procedure.
\newblock {\em Journal of the Royal Statistical Society: Series B (Statistical
  Methodology)\/}~{\em 64\/}(3), 499--517.

\bibitem[\protect\citeauthoryear{Genovese and Wasserman}{Genovese and
  Wasserman}{2004}]{GenoveseWasserman2004}
Genovese, C. and L.~Wasserman (2004, 06).
\newblock A stochastic process approach to false discovery control.
\newblock {\em Ann. Statist.\/}~{\em 32\/}(3), 1035--1061.

\bibitem[\protect\citeauthoryear{Goel and Rubin}{Goel and
  Rubin}{1977}]{GoelRubin1977}
Goel, P.~K. and H.~Rubin (1977).
\newblock On selecting a subset containing the best population-a bayesian
  approach.
\newblock {\em The Annals of Statistics\/}~{\em 5\/}(5), 969--983.

\bibitem[\protect\citeauthoryear{Gu and Koenker}{Gu and
  Koenker}{2017a}]{GuKoenker2017empirical}
Gu, J. and R.~Koenker (2017a).
\newblock Empirical bayesball remixed: Empirical bayes methods for longitudinal
  data.
\newblock {\em Journal of Applied Econometrics\/}~{\em 32\/}(3), 575--599.

\bibitem[\protect\citeauthoryear{Gu and Koenker}{Gu and
  Koenker}{2017b}]{GuKoenker2017unobserved}
Gu, J. and R.~Koenker (2017b).
\newblock Unobserved heterogeneity in income dynamics: An empirical bayes
  perspective.
\newblock {\em Journal of Business \& Economic Statistics\/}~{\em 35\/}(1),
  1--16.

\bibitem[\protect\citeauthoryear{Gu and Koenker}{Gu and
  Koenker}{2023}]{GuKoenker2020invidious}
Gu, J. and R.~Koenker (2023).
\newblock Invidious comparisons: Ranking and selection as compound decisions.
\newblock {\em Econometrica\/}~{\em 91\/}(1), 1--41.

\bibitem[\protect\citeauthoryear{Gupta}{Gupta}{1965}]{1965Gupta}
Gupta, S.~S. (1965).
\newblock On some multiple decision (selection and ranking) rules.
\newblock {\em Technometrics\/}~{\em 7\/}(2), 225--245.

\bibitem[\protect\citeauthoryear{Henderson and Newton}{Henderson and
  Newton}{2016}]{HendersonNewton2014}
Henderson, N.~C. and M.~A. Newton (2016).
\newblock Making the cut: improved ranking and selection for large-scale
  inference.
\newblock {\em Journal of the Royal Statistical Society. Series B, Statistical
  methodology\/}~{\em 78\/}(4), 781.

\bibitem[\protect\citeauthoryear{Jiang and Zhang}{Jiang and
  Zhang}{2009}]{jiang2009general}
Jiang, W. and C.-H. Zhang (2009).
\newblock General maximum likelihood empirical bayes estimation of normal
  means.
\newblock {\em The Annals of Statistics\/}~{\em 37\/}(4), 1647--1684.

\bibitem[\protect\citeauthoryear{Kamiński and Szufel}{Kamiński and
  Szufel}{2018}]{KaminskiPrzemyslaw2018}
Kamiński, B. and P.~Szufel (2018).
\newblock On parallel policies for ranking and selection problems.
\newblock {\em Journal of Applied Statistics\/}~{\em 45\/}(9), 1690--1713.

\bibitem[\protect\citeauthoryear{Koenker and Mizera}{Koenker and
  Mizera}{2014}]{koenker2014convex}
Koenker, R. and I.~Mizera (2014).
\newblock Convex optimization, shape constraints, compound decisions, and
  empirical bayes rules.
\newblock {\em Journal of the American Statistical Association\/}~{\em
  109\/}(506), 674--685.

\bibitem[\protect\citeauthoryear{Kwon and Zhao}{Kwon and
  Zhao}{2023}]{kwon2023f}
Kwon, Y. and Z.~Zhao (2023).
\newblock On f-modelling-based empirical bayes estimation of variances.
\newblock {\em Biometrika\/}~{\em 110\/}(1), 69--81.

\bibitem[\protect\citeauthoryear{Luo, Hong, Nelson, and Wu}{Luo
  et~al.}{2015}]{Luoetal2015}
Luo, J., L.~J. Hong, B.~L. Nelson, and Y.~Wu (2015).
\newblock Fully sequential procedures for large-scale ranking-and-selection
  problems in parallel computing environments.
\newblock {\em Operations Research\/}~{\em 63\/}(5), 1177--1194.

\bibitem[\protect\citeauthoryear{Mosteller}{Mosteller}{1948}]{1948Mostellar}
Mosteller, F. (1948).
\newblock {A $k$-Sample Slippage Test for an Extreme Population}.
\newblock {\em The Annals of Mathematical Statistics\/}~{\em 19\/}(1), 58 --
  65.

\bibitem[\protect\citeauthoryear{Ni, Ciocan, Henderson, and Hunter}{Ni
  et~al.}{2017}]{Nietal2017}
Ni, E.~C., D.~F. Ciocan, S.~G. Henderson, and S.~R. Hunter (2017).
\newblock Efficient ranking and selection in parallel computing environments.
\newblock {\em Operations Research\/}~{\em 65\/}(3), 821--836.

\bibitem[\protect\citeauthoryear{Panchapakesan}{Panchapakesan}{1971}]{Panchapakesan1971}
Panchapakesan, S. (1971).
\newblock On a subset selection procedure for the most probable event in a
  multinomial distribution**this research was supported in part by the office
  of naval research contract n00014-67-a-0226-00014 and the aerospace research
  laboratories contract af33 (615)67c1244 at purdue university. reproduction in
  whole or in part is permitted for any purposes of the united states
  government.
\newblock In S.~S. Gupta and J.~Yackel (Eds.), {\em Statistical Decision Theory
  and Related Topics}, pp.\  275--298. Academic Press.

\bibitem[\protect\citeauthoryear{Paulson}{Paulson}{1949}]{1949Paulson}
Paulson, E. (1949).
\newblock {A Multiple Decision Procedure for Certain Problems in the Analysis
  of Variance}.
\newblock {\em The Annals of Mathematical Statistics\/}~{\em 20\/}(1), 95 --
  98.

\bibitem[\protect\citeauthoryear{Silverman}{Silverman}{1986}]{Sil86}
Silverman, B.~W. (1986).
\newblock {\em Density estimation for statistics and data analysis}, Volume~26.
\newblock CRC press.

\bibitem[\protect\citeauthoryear{Storey}{Storey}{2002}]{Storey2002}
Storey, J.~D. (2002).
\newblock A direct approach to false discovery rates.
\newblock {\em Journal of the Royal Statistical Society: Series B (Statistical
  Methodology)\/}~{\em 64\/}(3), 479--498.

\bibitem[\protect\citeauthoryear{Sun and Cai}{Sun and
  Cai}{2007}]{sun2007oracle}
Sun, W. and T.~T. Cai (2007).
\newblock Oracle and adaptive compound decision rules for false discovery rate
  control.
\newblock {\em Journal of the American Statistical Association\/}~{\em
  102\/}(479), 901--912.

\bibitem[\protect\citeauthoryear{Sun and McLain}{Sun and
  McLain}{2012}]{SunMcLain:2012}
Sun, W. and A.~C. McLain (2012).
\newblock Multiple testing of composite null hypotheses in heteroscedastic
  models.
\newblock {\em Journal of the American Statistical Association\/}~{\em
  107\/}(498), 673--687.

\bibitem[\protect\citeauthoryear{Wand and Jones}{Wand and
  Jones}{1994}]{wand1996kernel}
Wand, M.~P. and M.~C. Jones (1994).
\newblock {\em Kernel Smoothing}, Volume~60 of {\em Chapman and Hall CRC
  Monographs on Statistics and Applied Probability}.
\newblock Chapman and Hall CRC.

\bibitem[\protect\citeauthoryear{Weinstein, Ma, Brown, and Zhang}{Weinstein
  et~al.}{2018}]{Weinsteinetal18}
Weinstein, A., Z.~Ma, L.~D. Brown, and C.-H. Zhang (2018).
\newblock Group-linear empirical bayes estimates for a heteroscedastic normal
  mean.
\newblock {\em Journal of the American Statistical Association\/}~{\em 0\/}(0),
  1--13.

\bibitem[\protect\citeauthoryear{Xie, Kou, and Brown}{Xie
  et~al.}{2012}]{Xieetal2012}
Xie, X., S.~Kou, and L.~D. Brown (2012).
\newblock Sure estimates for a heteroscedastic hierarchical model.
\newblock {\em Journal of the American Statistical Association\/}~{\em
  107\/}(500), 1465--1479.

\bibitem[\protect\citeauthoryear{Zhong, Liu, Luo, and Hong}{Zhong
  et~al.}{2022}]{Zhongetal2022}
Zhong, Y., S.~Liu, J.~Luo, and L.~J. Hong (2022).
\newblock Speeding up paulson’s procedure for large-scale problems using
  parallel computing.
\newblock {\em INFORMS Journal on Computing\/}~{\em 34\/}(1), 586--606.

\end{thebibliography}

	\newpage

\setcounter{page}{1} 

\appendix
\begin{center}\LARGE
	Online Supplementary Material for ``Ranking and Selection in Large-Scale Inference of Heteroscedastic Units''
\end{center}

\medskip

This Online Supplement contains the proofs of main theorems and propositions (\ref{sec:pf}), proofs of technical lemmas (\ref{lemmaproof}), a discussion on the grid size (\ref{grid}), a discussion on special cases of the r-value notion (\ref{sec:rv-pv-qv}), and a discussion on the ``nestedness" property (\ref{subsec:nest}).  
\section{Proof of main Theorems and Propositions}\label{sec:pf}

\subsection{Proof of Theorem \ref{thm:or}}

Observe that solving the constrained optimization problem \eqref{eq:RS-problem2} is equivalent to solving the subsequent problem:
$$
\mbox{maximize}\ \ E\left\{ \sum_i(X_i - \mu_0)\delta_i\right\} \ \ \ \mbox{subject to} \ \  E\left\{\sum_i\delta_i(\mbox{Clfdr}_i-\alpha)\right\}\leq 0. 
$$
We divide the discussion into the following scenarios. 

\noindent\textbf{1. Decisions for units in group 0.} 

Let $\pmb{\delta}$ be a decision rule satisfying $E\left\{\sum_i\delta_i(\mbox{Clfdr}_i-\alpha)\right\}\leq 0$. Denote by $\mathcal{R}(\pmb{\delta})$ the set of hypotheses rejected by $\pmb{\delta}$. Suppose that the null hypothesis $H_{0,j}$ from group 0 is not rejected by $\pmb{\delta}$. Consider another decision rule $\pmb{\delta}'$ with $\mathcal{R}(\pmb{\delta}')=\mathcal{R}(\pmb{\delta})\cup\{j\}$. It is clear that 
$$
\mbox{$E\left\{\sum_i\delta'_i(\mbox{Clfdr}_i-\alpha)\right\}\leq 0$ and $\sum_i(x_i - \mu_0)\delta'_i \geq \sum_i(x_i - \mu_0)\delta_i.$}
$$ 
Hence, the optimal procedure must reject all hypotheses from group 0.

\noindent\textbf{2. Decisions for units in group 3.} 

Next, suppose $\pmb{\delta}$ rejects the null hypothesis $H_{0,j}$ from group 3. Consider a new decision rule $\pmb{\delta}'$ with $\mathcal{R}(\pmb{\delta}')=\mathcal{R}(\pmb{\delta})\backslash\{j\}$. It is  clear that 
$$\mbox{$E\left\{\sum_i\delta'_i(\mbox{Clfdr}_i-\alpha)\right\}\leq 0$ and $\sum_i(x_i - \mu_0)\delta'_i > \sum_i(x_i - \mu_0)\delta_i$.}$$ 
Hence, the optimal procedure does not reject any hypothesis from group 3.

\noindent\textbf{3. Decisions for units in group 1 and group 2.}  

Let $\mathcal{R}_{\pmb{\delta}}^+=\{i \in \mathcal{R}(\pmb{\delta}): \mbox{Clfdr}_i-\alpha>0 \}$, and $\mathcal{R}_{\pmb{\delta}}^-=\{i \in \mathcal{R}(\pmb{\delta}): \mbox{Clfdr}_i-\alpha\leq 0 \}$. Then $\mathcal{R}_{\pmb{\delta}}^+$ and $\mathcal{R}_{\pmb{\delta}}^+$ respectively correspond to the decisions for units in group 1 and group 2. 

\begin{remark}
	We pause momentarily to offer clarification on the key concepts that will be presented in the remainder of the proof. It is important to note that the $\alpha$-investing and $\mu$-investing processes are interdependent, which means that the optimal cutoff in group 1 is contingent on the cutoff chosen in group 2. As a result, the derivation of the optimal decision rule can be challenging. However, an important observation is that if any decision procedure deviates from the oracle rule for group 1, it can be uniformly enhanced by ranking hypotheses in group 1 based on the ordering of $T_i$ in descending order and then selecting a suitable threshold. This argument applies similarly in the opposite direction for selection of units in group 2. Therefore, although the process of determining the optimal pairs of $(c_1, c_2)$ may be complex, the format of the optimal decision rule can be determined. 
\end{remark}

Subsequently, we will demonstrate separately that both $\mathcal{R}_{\pmb{\delta}_{OR}}^+$ and $\mathcal{R}_{\pmb{\delta}_{OR}}^-$, when holding the part fixed, correspond to optimal rejection sets that cannot be further improved.

Suppose $\pmb{\delta}$ satisfies $E\left\{\sum_i\delta_i(\mbox{Clfdr}_i-\alpha)\right\}\leq 0$ and $\mathcal{R}_{\pmb{\delta}^{OR}}^-=\mathcal{R}_{\pmb{\delta}}^-$. The oracle rule on group 1 can be expressed as
\beq
\delta^{OR}_i = 
\begin{cases}
	0 & \mbox{if } X_i- \mu_0 \leq \xi^{-1}(c_1^{OR}) (\mbox{Clfdr}_i - \alpha) \\
	1 & \mbox{if } X_i- \mu_0 > \xi^{-1}(c_1^{OR}) (\mbox{Clfdr}_i - \alpha). \\
	
\end{cases}
\eeq
Let $\mathcal{I}^+=\{ i:E(\delta^{OR}_i-\delta_i )>0 \}$ and $\mathcal{I}^-=\{ i:E(\delta^{OR}_i-\delta_i ) < 0 \}$. For $i \in \mathcal{I}^+$, we have $\delta^{OR}_i=1$ and hence $ X_i - \mu_0 > \xi^{-1}(c_1^{OR}) (\mbox{Clfdr}_i - \alpha)$. Similarly for $i \in \mathcal{I}^-$, we have $\delta^{OR}_i=0$ and $ X_i - \mu_0 < \xi^{-1}(c_1^{OR}) (\mbox{Clfdr}_i - \alpha)$. Thus,
\begin{equation}\label{pf1}
	\sum_{i\in \mathcal{I}^+\cup \mathcal{I}^-}E\{  \delta^{OR}_i-\delta_i  \}\{ (X_i-\mu_{0})-\xi^{-1}(c_1^{OR})(\mbox{Clfdr}_i-\alpha)  \}\geq 0.
\end{equation}

Given $\mathcal{R}_{\pmb{\delta}^{OR}}^-=\mathcal{R}_{\pmb{\delta}}^-$, $c^{OR}_1$ is chosen as small as possible such that $E\left\{\sum_i\delta^{OR}_i(\mbox{Clfdr}_i-\alpha)\right\}= 0$ or until the entire group 1 is rejected. 
Such $c_1^{OR}$ exists because $E\left\{\sum_i\delta^{OR}_i(\mbox{Clfdr}_i-\alpha)\right\}$ is a continuous and monotone function of $c_1^{OR}$ when ignoring the decisions in the other group. The argument follows from that in \citep{CaiSunWang2019}. 
In particular, this implies 
\begin{equation}\label{pf2}
	E\left\{\sum_i\delta_i(\mbox{Clfdr}_i-\alpha)\right\}\leq E\left\{\sum_i\delta^{OR}_i(\mbox{Clfdr}_i-\alpha)\right\}.
\end{equation}
Recall the definition of the power function
$$
ETP^*_{\pmb{\delta}}=E\left\{ \sum_{i=1}^{m}(X_i-\mu_0)\delta_i     \right\}.
$$
Using (\ref{pf1}) and (\ref{pf2}), we conclude that $ETP^*_{\pmb{\delta}^{OR}}\geq ETP^*_{\pmb{\delta}}$.

By employing a similar line of reasoning as described above, it can be shown that the optimal procedure involves ranking hypotheses in group 2 in ascending order of $T_i$ and selecting an appropriate threshold.

Finally, we combine the claims from the four groups, and claim that the oracle rule is optimal in the sense of \eqref{eq:RS-problem2}.  

\subsection{Proof of Proposition \ref{monotone}}

It is worth noting that if $(a,b)$ and $(c,d)$ are points on $L$ and $a\geq c$, then we must have $c\leq d$. Additionally, if $b\leq c$, then we must have $\mbox{ETP}^*_{\pmb{\delta}(a,b)}\geq \mbox{ETP}^*_{\pmb{\delta}(a,c)}$.


Suppose $\mbox{mFDR}_{\pmb{\delta}}> \mbox{mFDR}_{\pmb{\delta}'}$, the we can find another point $(r'_1,\tilde{r}'_2)\in l$ such that $\tilde{r}'_2\in [r_2,r'_2)$, $\mbox{mFDR}_{\pmb{\delta}}= \mbox{mFDR}_{\pmb{\delta}(r'_1,\tilde{r}'_2)}$ and $\mbox{ETP}^*_{\pmb{\delta}'} \leq \mbox{ETP}^*_{\pmb{\delta}(r'_1,\tilde{r}'_2)} \leq \mbox{ETP}^*_{\pmb{\delta}} $. Similarly, suppose $\mbox{mFDR}_{\pmb{\delta}'}> \mbox{mFDR}_{\pmb{\delta}''}$, the we can find another point $(c''_1,\tilde{c}''_2)\in l$ such that $\tilde{c}''_2\in [c'_2,c''_2)$, $\mbox{mFDR}_{\pmb{\delta}'}= \mbox{mFDR}_{\pmb{\delta}(c''_1,\tilde{c}''_2)}$ and 
$$
\mbox{ETP}^*_{\pmb{\delta}''} \leq \mbox{ETP}^*_{\pmb{\delta}(c''_1,\tilde{c}''_2)} \leq \mbox{ETP}^*_{\pmb{\delta}'}.
$$ 
Thus, if we can show the claim holds under the assumption that $\mbox{mFDR}_{\pmb{\delta}}= \mbox{mFDR}_{\pmb{\delta}'}=\mbox{mFDR}_{\pmb{\delta}''}$, then the desired result follows. 


We introduce some notations:
\begin{itemize}
	\item $\pmb{\mbox{CLfdr}}=(\mbox{CLfdr}_1,\mbox{CLfdr}_2,...,\mbox{CLfdr}_m)$.
	\item $\pmb{T}=(T_1, T_2,...,T_m)$.
	\item $\pmb{x}_{I_k}$ is the vector $\pmb{x}$ restricted to group $k$.
	\item $\pmb{x}|_{\pmb{y}}$ is the vector $\pmb{x}$ restricted to the non-zero entries in $\pmb{y}$.
	\item $\mathbbm{1}=(1,1,..,1)$ a vector of $1$'s. 
	\item $\mbox{ave}(\pmb{x}/\pmb{y})=\pmb{x}\mathbbm{1}^t/\pmb{y}\mathbbm{1}^t.$
	\item if $\pmb{a}=(a_1,\ldots,a_n)$ is a vector and $b$ is a number, then $\pmb{a}-b=(a_1-b,\ldots,a_n-b)$.
\end{itemize}

By our ranking strategy for the units in group 1 and group 2 in the oracle rule, we have 
\begin{eqnarray} \label{eq2}
	\mbox{ave}\{  (\pmb{x}-\mu_0)_{I_2}|_{\pmb{\delta}''-\pmb{\delta}'}/(\pmb{\mbox{CLfdr}}-\alpha)_{I_2}|_  {\pmb{\delta}''-\pmb{\delta}'}  \} & \geq &  \mbox{ave}\{  (\pmb{x}-\mu_0)_{I_2}|_{\pmb{\delta}'-\pmb{\delta}}/(\pmb{\mbox{CLfdr}}-\alpha)_{I_2}|_  {\pmb{\delta}'-\pmb{\delta}}  \},   \\ \label{eq3}
	\mbox{ave}\{  (\pmb{x}-\mu_0)_{I_1}|_{\pmb{\delta}''-\pmb{\delta}'}/(\pmb{\mbox{CLfdr}}-\alpha)_{I_1}|_  {\pmb{\delta}''-\pmb{\delta}'}  \} & \leq & \mbox{ave}\{  (\pmb{x}-\mu_0)_{I_1}|_{\pmb{\delta}'-\pmb{\delta}}/(\pmb{\mbox{CLfdr}}-\alpha)_{I_1}|_  {\pmb{\delta}'-\pmb{\delta}}  \}. 
\end{eqnarray}

Next, note that $\mbox{ETP}^*_{\pmb{\delta}}\geq \mbox{ETP}^*_{\pmb{\delta}'}$ implies 
$$
(\pmb{\delta}'-\pmb{\delta})_{I_1} (\pmb{x}-\mu_0\mathbbm
{1}) ^t_{I_1} \leq -(\pmb{\delta}'-\pmb{\delta})_{I_2} (E(\pmb{\mu})-\mu_0\mathbbm
{1})^t _{I_2},
$$
which can be re-written as
\begin{align}\label{pf3}
	&\mbox{ave}\{  (\pmb{x}-\mu_0)_{I_1}|_{\pmb{\delta}'-\pmb{\delta}}/(\pmb{\mbox{CLfdr}}-\alpha)_{I_1}|_  {\pmb{\delta}'-\pmb{\delta}}  \}\times  (\pmb{\delta}'-\pmb{\delta})_{I_1}(\pmb{\mbox{CLfdr}}-\alpha \mathbbm{1})^t_{I_1}\\ \nonumber
	\leq &- \mbox{ave}\{  (\pmb{x}-\mu_0)_{I_2}|_{\pmb{\delta}'-\pmb{\delta}}/(\pmb{\mbox{CLfdr}}-\alpha)_{I_2}|_  {\pmb{\delta}'-\pmb{\delta}}  \}\times  (\pmb{\delta}'-\pmb{\delta})_{I_2}(\pmb{\mbox{CLfdr}}-\alpha \mathbbm{1})^t_{I_2}.
\end{align}

The condition $\mbox{mFDR}_{\pmb{\delta}}= \mbox{mFDR}_{\pmb{\delta}'}$ implies that
$$
(\pmb{\delta}'-\pmb{\delta})_{I_1}(\pmb{\mbox{CLfdr}}-\alpha \mathbbm{1})^t_{I_1}= -(\pmb{\delta}'-\pmb{\delta})_{I_2}(\pmb{\mbox{CLfdr}}-\alpha \mathbbm{1})^t_{I_2}.
$$
According to (\ref{pf3}), we have
$$
\mbox{ave}\{  (\pmb{x}-\mu_0)_{I_2}|_{\pmb{\delta}'-\pmb{\delta}}/(\pmb{\mbox{CLfdr}}-\alpha)_{I_2}|_  {\pmb{\delta}'-\pmb{\delta}}  \}\geq  \mbox{ave}\{  (\pmb{x}-\mu_0)_{I_1}|_{\pmb{\delta}'-\pmb{\delta}}/(\pmb{\mbox{CLfdr}}-\alpha)_{I_1}|_  {\pmb{\delta}'-\pmb{\delta}}  \}.
$$
Combining (\ref{eq2}) with (\ref{eq3}), we have
$$
\mbox{ave}\{  (\pmb{x}-\mu_0)_{I_2}|_{\pmb{\delta}''-\pmb{\delta}'}/(\pmb{\mbox{CLfdr}}-\alpha)_{I_2}|_  {\pmb{\delta}''-\pmb{\delta}'}  \}\geq  \mbox{ave}\{  (\pmb{x}-\mu_0)_{I_1}|_{\pmb{\delta}''-\pmb{\delta}'}/(\pmb{\mbox{CLfdr}}-\alpha)_{I_1}|_  {\pmb{\delta}''-\pmb{\delta}'}  \}.
$$

Similarly the condition $\mbox{mFDR}^*_{\pmb{\delta}''}= \mbox{mFDR}^*_{\pmb{\delta}'}$ implies that 
$$
(\pmb{\delta}''-\pmb{\delta}')_{I_1}(\pmb{\mbox{CLfdr}}-\alpha \mathbbm{1})^t_{I_1}= (\pmb{\delta}''-\pmb{\delta}')_{I_2}(\pmb{\mbox{CLfdr}}-\alpha \mathbbm{1})^t_{I_2}.
$$

Hence we have
\begin{align*}
	&(\pmb{\delta}''-\pmb{\delta}')_{I_2} (\pmb{x}-\mu_0\mathbbm
	{1}) ^t_{I_2}\\
	=&\mbox{ave}\{  (\pmb{x}-\mu_0)_{I_2}|_{\pmb{\delta}''-\pmb{\delta}'}/(\pmb{\mbox{CLfdr}}-\alpha)_{I_2}|_  {\pmb{\delta}''-\pmb{\delta}'}  \}\times   (\pmb{\delta}''-\pmb{\delta}')_{I_2}(\pmb{\mbox{CLfdr}}-\alpha \mathbbm{1})^t_{I_2}\\
	\leq &-\mbox{ave}\{  (\pmb{x}-\mu_0)_{I_1}|_{\pmb{\delta}''-\pmb{\delta}'}/(\pmb{\mbox{CLfdr}}-\alpha)_{I_1}|_  {\pmb{\delta}''-\pmb{\delta}'}  \}\times  (\pmb{\delta}''-\pmb{\delta}')_{I_1}(\pmb{\mbox{CLfdr}}-\alpha \mathbbm{1})^t_{I_1}\\
	=&-(\pmb{\delta}''-\pmb{\delta}')_{I_1} (\pmb{x}-\mu_0\mathbbm
	{1}) ^t_{I_1}.
\end{align*}

Now we can see that
$$
(\pmb{\delta}''-\pmb{\delta}')_{I_2} (\pmb{x}-\mu_0\mathbbm
{1}) ^t_{I_2}+(\pmb{\delta}''-\pmb{\delta}')_{I_1} (\pmb{x}-\mu_0\mathbbm
{1}) ^t_{I_1}\leq 0,$$
which implies that $\mbox{ETP}^*_{\pmb{\delta}'}\geq \mbox{ETP}^*_{\pmb{\delta}''}$ and the desired result follows.


\subsection{Proof of Proposition \ref{prop1}}
We first state a useful lemma:
\begin{lemma}\label{lem3}
	Suppose $\mu_i\overset{iid}{\sim} g(\cdot)$, for $i=1,..., m$. Let $\hat{g}$ be the empirical density function $\sum_{i=1}^{n}\delta_{\mu_i}(\cdot)$. Let $f(x)=\int_{-\infty}^{\infty}\phi_{\sigma}(\mu-x)g(\mu)d\mu$ and $\hat{f}(x)=\int_{-\infty}^{\infty}\phi_{\sigma}(\mu-x)\hat{g}(\mu)d\mu$. Then for every $x$, $E_{\pmb{\mu}}|f(x)-\hat{f}(x) |^2\rightarrow 0$ as $m \rightarrow \infty$. 
\end{lemma}

Lemma \ref{lem3} implies it is possible to find a set $\{\mu_1,\ldots,\mu_m\}$ and $\hat{f}_\sigma(x)=\frac{1}{m}\sum_{i=1}^{m}\phi_\sigma(x-\mu_i)$ such that for all $x$, $|f_{\sigma}(x)-\hat{f}_{\sigma}(x) |^2\rightarrow 0$. 
Consider the following set of functions 
$$
\left\{  \sum_{i=0}^{k-1}w_i\phi_{\sigma}(x-s-i\eta) | \sum_{i=0}^{k-1}w_i=1, w_i\geq 0 \ \ \forall i \right\}.
$$ 
We can make the grid fine enough so that for any $\epsilon>0$ and $i$, there exists $s_i\in \{s,s+\eta,...,s+(k-1)\eta\}$ such that $|\mu_i-s_i|<\epsilon$. Hence 
\begin{align*}
	\big|\frac{1}{m}\sum_{i=1}^{m}\phi_\sigma(x-\mu_i)-\dfrac{1}{m}\sum_{i=1}^{m}\phi_\sigma(x-s_i)\big|^2&=\dfrac{1}{m^2}\big|\sum_{i=1}^{m}\phi_\sigma(x-\mu_i)-\sum_{i=1}^{m}\phi_\sigma(x-s_i)\big|^2\\
	&\leq \dfrac{1}{m}\sum_{i=1}^{m}| \phi_{\sigma}(x-s_j)-\phi_{\sigma}(x-\mu_i) |^2.
\end{align*}
If we let $\epsilon\rightarrow 0$, then $| \phi_{\sigma}(x-s_i)-\phi_{\sigma}(x-\mu_i) |^2\rightarrow 0$ for $i=1,...,m$. It follows that there exists $$\psi_{\sigma}\in \{  \sum_{i=0}^{k-1}w_i\phi_{\sigma}(x-s-i\eta) | \sum_{i=0}^{k-1}w_i=1, w_i\geq 0 \ \ \forall i \}$$ such that 
$ | {f}_\sigma(x)-\psi_\sigma(x) |^2\rightarrow 0$. 

Using standard arguments
in density estimation theory (e.g. \cite{wand1996kernel}), we have 
$$
\mathbb{E}\|\hat{f}^m_\sigma-f_\sigma\|_2^2= O\{(mh_xh_\sigma)^{-1}+h^4_x+h^4_\sigma\}. 
$$
By assumption (A2) $(mh_xh_\sigma)^{-1}+h^4_x+h^4_\sigma\rightarrow 0$.
It follows that 
$$
\frac{1}{m}\sum_{i=1}^{m}\{\hat{f}_i(x_i)-\hat{f}^m_i(x_i)\}^2\overset{p}{\rightarrow}\frac{1}{m}\sum_{i=1}^{m}\{\hat{f}_i(x_i)-f_{\sigma_i}(x_i)\}^2. 
$$
By definition of the minimization problem, we have $\frac{1}{m}\sum_{i=1}^{m}\{\hat{f}_i(x_i)-f_{\sigma_i}(x_i)\}^2\overset{p}{\rightarrow}0$.
Thus,  
$E_{\pmb{x},\pmb{\sigma}} E_{\sigma,x}| \hat{f}_\sigma(x)-f_\sigma(x) |^2\rightarrow 0$ and $E_{\pmb{x},\pmb{\sigma}} E_{\sigma,x}| \hat{f}_{0,\sigma}(x)-f_{0,\sigma}(x) |^2\rightarrow 0$, where $\hat{f}_{0,\sigma}(x)=\sum_{s_i<\mu_0}w_i\phi_\sigma(x-s_i)$ and $f_{0,\sigma}(x)=\int_{-\infty}^{\mu_0}\phi_{\sigma}(x-\mu)g(\mu)d\mu$. Here $E_{\sigma,x}$ is taken with respect to $\sigma$ and $x$, $E_{\pmb{x},\pmb{\sigma}}$ is taken with respect to the data that are used to construct $\hat{f}$.

Note that $f_\sigma$ is continuous, then there exists $K_1=[-M,M]$ such that $P(x\in K^c_1)\rightarrow0$ as $M\rightarrow\infty$. Let $\inf_{x\in K_1}f_\sigma(x)=l_0$ and $A_{l_0}=\{x:|\hat{f}_\sigma(x)-f_\sigma(x)|\geq l_0/2\}$. Since 
$$
E_{\pmb{x},\pmb{\sigma}} E_{\sigma,x}| \hat{f}_{0,\sigma}(x)-f_{0,\sigma}(x) |^2\geq (l_0/2)^2P(A^f_\epsilon),
$$
it follows that $P(A_{l_0})\rightarrow 0$. Thus $\hat{f}_\sigma$ and $f_\sigma$ are bounded below by a positive number for large $n,m$ except for an event that has a low probability. Similar arguments can be applied to the upper bound of $\hat{f}_\sigma$ and $f_\sigma$, as well as to the upper and lower bounds for $\hat{f}_{0,\sigma}$ and $f_{0,\sigma}$. Therefore, we conclude that $\hat{f}_{0,\sigma}$, $\hat{f}_{\sigma}$, $f_{0,\sigma}$ and $f_{\sigma}$. are all bounded in the interval $[l_a, l_b]$, $0<l_a<l_b<\infty$ for large $n,m$ except for an event, say $A_{l_0}$ that has low probability. 
Let $\hlfdr(x,\sigma)=\hat{f}_{0,\sigma}(x)/\hat{f}_\sigma(x)$ and $\mbox{Clfdr}(x,\sigma)=f_{0,\sigma}(x)/f_\sigma(x)$. We have 
$$\hlfdr(x,\sigma)-\mbox{Clfdr}(x,\sigma)=\frac{\hat{f}_{0,\sigma}(x)f_\sigma(x)  -f_{0,\sigma}(x)\hat{f}_\sigma(x)}{\hat{f}_\sigma(x)f_\sigma(x)}.$$  
Since $| \hlfdr-\mbox{Clfdr}   |^2$ is bounded by 1, we have
\begin{eqnarray*}
	&& E_{\pmb{x},\pmb{\sigma}}E_{\sigma,x}\{\hlfdr(x,\sigma)-\mbox{Clfdr}(x,\sigma)\}^2 \\ 
	& \leq &  P(A_{l_0})+c_1E_{\pmb{x},\pmb{\sigma}}E_{\sigma,x}\{ \hat{f}_{0,\sigma}(x)-f_{0,\sigma}(x)   \}^2+E_{\pmb{x},\pmb{\sigma}}E_{\sigma,x}\{\hat{f}_\sigma(x)-f_\sigma(x) \}^2.
\end{eqnarray*}
Thus, $E_{\pmb{x},\pmb{\sigma}}E_{\sigma,x}\{\hlfdr(x,\sigma)-\mbox{Clfdr}(x,\sigma)\}^2\rightarrow 0.$ Let $B_\delta=\{x,\sigma: |\hlfdr(x,\sigma)|- \mbox{Clfdr}(x,\sigma)|>\delta \}$. Then we have 
$$\delta^2P(B_\delta)\leq E_{\pmb{x},\pmb{\sigma}}E_{\sigma,x}\{\hlfdr(x,\sigma)-\mbox{Clfdr}(x,\sigma)\}^2\rightarrow 0,$$ and the desired result follows.

\subsection{Proof of Theorem \ref{dd}}
We begin with a summary of notation used throughout the proof:

\begin{itemize}
	\item $Q_m(c_1,c_2) = m^{-1}\sum_{i=1}^m (\mbox{Clfdr}_i-\alpha) \pmb{\delta}(c_1,c_2)(S_i)$. 
	\item $\hQ_m(c_1,c_2) = m^{-1}\sum_{i=1}^m (\hlfdr_i-\alpha) \pmb{\delta}^{}(c_1,c_2)(\hat{S}_i)$. 
	\item $Q_{\infty}(c_1,c_2) = E\{(\mbox{Clfdr}-\alpha)\pmb{\delta}(c_1,c_2)(S)\}$.
	\item $\hat{H}_m(c_1,c_2) = m^{-1}\sum_{i=1}^m (X_i-\mu_{0}) \pmb{\delta}^{}(c_1,c_2)(\hat{S}_i)$. 
	\item $H_{\infty}(c_1,c_2) = E\{m^{-1}\sum_{i=1}^m (X_i-\mu_{0}) \pmb{\delta}^{}(c_1,c_2)(S_i)\}$. 
	\item 	$(c^{OR}_1,c^{OR}_2)=\argmax_{(c_1,c_2):Q_\infty(c_1,c_2)\leq 0}\left\{ H_{\infty}(c_1,c_2)   \right\}.$
	\item $(\hat{c}_1,\hat{c}_2)=\argmax_{(c_1,c_2):\hat{Q}(c_1,c_2)\leq 0}\left\{ \hat{H}(c_1,c_2)   \right\}. $
\end{itemize}
We first show $\hQ_m(c_1.c_2) \overset{p}\rightarrow Q_{\infty}(c_1,c_2)$. Note that $Q_m(c_1,c_2)\overset{p}{\rightarrow}Q_{\infty}(c_1,c_2)$ by the WLLN, so that we only need to establish $\hQ_m(c_1,c_2)\overset{p}{\rightarrow} Q_m(c_1,c_2)$. We state a lemma that will be useful for the proof:
\begin{lemma}\label{l2conv}
	Let $U_i=(\mbox{Clfdr}_i-\alpha)\delta(c_1,c_2)(S_i)$ and $\hat{U}_i=(\hlfdr_i-\alpha)\delta(c_1,c_2)(\hat{S}_i)$ then $E(U_i-\hat{U}_i)^2=o(1)$.
\end{lemma}

By Lemma \ref{l2conv} and Cauchy-Schwartz inequality, we have 
$$
E\left\{\left(\hat{U}_i-U_i\right)\left(\hat{U}_j-U_j\right)\right\} = o(1).
$$ 
Let $L_m = \sum_{i=1}^m \left(\hat{U}_i - U_i\right)$. 
It follows that 
$$
\mbox{Var}\left( m^{-1} L_m  \right) \leq m^{-2}\sum_{i=1}^{m}  E\left\{ \left( \hat{U}_i - U_i\right)^2 \right\} +O\left(\frac{1}{m^2}\sum_{i,j:i\neq j}  E\left\{\left(\hat{U}_i-U_i\right)\left(\hat{U}_j-U_j\right)\right\}\right)= o(1).
$$
By Lemma \ref{l2conv}, $ E(m^{-1}L_m)\rightarrow 0$, applying Chebyshev's inequality, we obtain 
$$
m^{-1}L_m = \hQ(c_1,c_2) - Q(c_1,c_2) \overset{p}\rightarrow 0.
$$ 

Next we show $\hQ_m(c_1.c_2) \rightarrow Q_{\infty}(c_1,c_2)$ uniformly.
Since $\hQ_m(c_1,c_2) \overset{p}\rightarrow Q_{\infty}(c_1,c_2)$ for all $(c_1,c_2)$ on the rectangle $\left[  c^-_1 ,c^+_1\right] \times \left[ c^-_2 , c^+_2\right] $, given any $\epsilon>0$ the Lebesgue measure of the set 
$$
\{(c_1,c_2)\in \left[  c^-_1 ,c^+_1\right] \times \left[ c^-_2 , c^+_2\right] : |\hQ_m(c_1,c_2)- Q_{\infty}(c_1,c_2)|>\epsilon \}
$$ 
approaches 0. Suppose there exists $\epsilon>0$ such that for any $M$ there is $m>M$ and $(c_1,c_2)$ with $\hQ_m(c_1,c_2) - Q_{\infty}(c_1,c_2) >2\epsilon$, since $Q_{\infty}$ is smooth, there exists a square $S_\delta(c_1,c_2)$ centered at $(c_1,c_2)$ with side length $\delta$ such that 
$$
\hQ_m(c_1,c_2) - Q_{\infty}(c'_1,c'_2) >\epsilon, \ \forall (c'_1,c'_2)\in S_\delta(c_1,c_2).
$$ 
Consider the triangle with vertices at $\{(c_1,c_2),\ (c_1-\delta/2,c_2),(c_1,c_2-\delta/2)\}$, call this triangle $\Delta$. It is clear that by definition of $\hat{Q}$ we have
$$
\min\{  \hQ_m(c_1,c_2),\hQ_m(c_1-\delta/2,c_2),\hQ_m(c_1,c_2-\delta/2)  \}\leq \inf_{(a,b)\in \Delta}\hQ_m(a,b),
$$
$$
\max\{  \hQ_m(c_1,c_2),\hQ_m(c_1-\delta/2,c_2),\hQ_m(c_1,c_2-\delta/2)  \}\geq \sup_{(a,b)\in \Delta}\hQ_m(a,b).
$$
Since $\hQ_m(c_1-\delta/2,c_2)\geq \hQ_m(c_1,c_2)$ and $\hQ_m(c_1,c_2-\delta/2) \geq \hQ_m(c_1,c_2)$. It follows that $\hQ_m(a,b) - Q_{\infty}(a,b) >\epsilon, \ \forall (a,b)\in \Delta $. Note that $\delta$ only depends on $\epsilon$, hence the area of $\Delta$ does not go to 0 as $m\rightarrow \infty$, a contradiction. Similarly, there is no $(c_1,c_2)$ and $\epsilon>0$ such that for any $M$, there exists $m>M$ with 
$$
\hQ_m(c_1,c_2) - Q_{\infty}(c_1,c_2) <-2\epsilon. 
$$
This implies $\hQ_m(c_1,c_2) \rightarrow Q_{\infty}(c_1,c_2)$ uniformly. Use similar arguments, we can also show $\hat{H}_m(c_1,c_2) \rightarrow H_{\infty}(c_1,c_2)$ uniformly. Define
$$
\hat{l}=\left\{   (c_1,\hat{S}_i): \hat{S}_i\in \mbox{group 2} , c_1=\max\{k: \hat{S}^{(k)}\in \mbox{group 1 and } \sum_{j=1}^{m}(\widehat{\mbox{Clfdr}}_i-\alpha)\delta(\hat{S}^{(k)}, \hat{S}_i) (\hat{S}_j)\leq 0\} \right\}.
$$
It is clear that the data-driven algorithm only searches among the points on $\hat{l}$. By uniform convergence, given any $\epsilon>0$, we can find $M$ such that for all $m>M$ $|\hat{Q}_m(c_1,c_2) \rightarrow Q_{\infty}(c_1,c_2)|<\epsilon$ for all $(c_1,c_2)\in \hat{l}$. This shows $\mbox{mFDR}_{\pmb{\delta}^{DD}}=\alpha+o(1)$.

Next we show $\pmb{\delta}^{DD}$ is asymptotically optimal. By uniform continuity of $H_\infty $, given any $\epsilon>0$, there exists $\delta>0$ such that $|H_\infty(a,b)-H_\infty(c,d)|<\epsilon$ for all $\|(a,b)-(c,d)\|\leq \delta$. With probability goes to 1, there exists a point $(a,b)\in D_\delta(c_1^{OR},c_2^{OR})$. By uniform convergence of $\hat{H}_m$ to $H_\infty$, we can choose $m$ big enough so that $|\hat{H}_m(a,b)-H_\infty(a,b)|<\epsilon$,  thus $$|\hat{H}(a,b)-H_\infty(c^{OR}_1,c^{OR}_2) |\leq |\hat{H}_m(a,b)-H_\infty(a,b)|+ |H_\infty(a,b)-H_\infty(c^{OR}_1,c^{OR}_2)|< 2\epsilon.$$

Again by uniform convergence we have $|\hat{H}_m(\hat{c}_1,\hat{c}_2) -H_\infty(\hat{c}_1,\hat{c}_2)|<\epsilon$ for all $m$ big enough. By definition $\hat{H}_m(\hat{c}_1,\hat{c}_2)>\hat{H}_m(a,b)$, thus 
$$
H_\infty(\hat{c}_1,\hat{c}_2)>\hat{H}_m(a,b)-\epsilon> H_\infty(c^{OR}_1,c^{OR}_2)-3\epsilon. 
$$
It follows that $\mbox{ETP}^*_{\pmb{\delta}^{DD}}/\mbox{ETP}^*_{\pmb{\delta}^{OR}}\geq 1+o(1)$, proving the desired result.

\subsection{Proof of Theorem \ref{consistent}}
Since the two definitions of r-value share the same selection procedure, it suffices 
to show that if $X_i>X_j$ and $\text{Clfdr}_i< \text{Clfdr}_j$ then the rejection of hypothesis $j$ implies the rejection of hypothesis $i$. 
We break the proof into several cases:\\
\textit{case 1}: If hypothesis $j$ belongs to group 0, then by definition hypothesis $i$ also belongs to group 0, hence it is also rejected.\\
\textit{case 2}: If hypothesis $j$ belongs to group 1, then hypothesis $i$ is either in group 0 or in group 1 with $T_i>T_j$. By definition of the oracle procedure, hypothesis $i$ is rejected.\\
\textit{case 3}: If hypothesis $j$ belongs to group 2, then hypothesis $i$ is either in group 0 or in group 2 with $T_i<T_j$. By definition of the oracle procedure, hypothesis $i$ is rejected.

The proof for the data-driven procedure with $\text{Clfdr}$ substituted by $\widehat{\mbox{Clfdr}}$ and $T$ replaced by $\hat{T}$ can be derived using the same argument.

\section{Proof of Lemmas}\label{lemmaproof}

\subsection{Proof of Lemma \ref{lem3}}

We use the bias-variance decomposition:
$$
E\{f(x)-\hat{f}(x) \}^2=\{E\hat{f}(x)-f(x)\}^2+\mbox{Var}\hat{f}(x).
$$
Write $\hat{g}=\sum_{i=1}^{m}\frac{1}{m}\delta_{\mu_i}(\cdot)$ as a mixture of point mass where $\mu_i\overset{iid}{\sim} g$. By definition,
$$
E\hat{f}(x)=E\sum_{i=1}^{m}\dfrac{1}{m}\phi_{\sigma}(x-\mu_i)=E\phi_\sigma(x-\mu)=\int_{-\infty}^{\infty}\phi_\sigma(x-\mu)g(\mu)d\mu=f(x).
$$
$\{E\hat{f}(x)-f(x)\}^2=0$. Also since $\phi$ is bounded, it follows that $\mbox{Var}\{\phi_\sigma(x-\mu_i)\}<\infty$. Therefore 
\begin{align*}
	\mbox{Var}\hat{f}(x)&=\mbox{Var}\left\{\int_{-\infty}^{\infty}\phi_{\sigma}(\mu-x)\hat{g}(\mu)d\mu\right\}\\
	&=\mbox{Var}\left\{ \dfrac{1}{m} \sum_{i=1}^{m}\phi_\sigma(x-\mu_i)      \right\}\\
	&=\dfrac{1}{m}\mbox{Var}\{\phi_\sigma(x-\mu_i)\}\rightarrow 0.
\end{align*}

\subsection{Proof of Lemma \ref{l2conv}}
We state a fact that will be helpful: 
\begin{lemma}\label{rconv}
	$\hat{S}_i\overset{P}{\rightarrow}S_i$.
\end{lemma}
Lemma \ref{rconv} is proved in  section \ref{pfrconv}.
By definition of $U_i$ and $\hat{U}_i$ we have the following:
\begin{align*}
	(U_i-\hat{U}_i)^2&=(\mbox{Clfdr}_i-\hlfdr_i)^2\mathbb{I}\{\delta(c_1,c_2)(S_i)=\delta(c_1,c_2)(\hat{S}_i)=1\}\\
	&+(\mbox{Clfdr}_i-\alpha)^2\mathbb{I}\{ \delta(c_1,c_2)(S_i)=1,\delta(c_1,c_2)(\hat{S}_i)= 0 \}\\
	&+(\hlfdr_i-\alpha)^2\mathbb{I}\{ \delta(c_1,c_2)(S_i)=0,\delta(c_1,c_2)(\hat{S}_i)= 1 \}
\end{align*}
Denote the three sums on the RHS as $I$, $II$, and $III$ respectively. By Proposition \ref{prop1}, $E(I) = o(1)$. To show $E(II+III)=o(1)$ we only need to show $P\{ \delta(c_1,c_2)(S_i)\neq \delta(c_1,c_2)(\hat{S}_i) \}=o(1)$. 
We say $S_i$ or $\hat{S}_i$ is from group $a$ if  $(X_i,\text{Clfdr}_i)$ is from group $a$. 
$\delta(c_1,c_2)(S_i)\neq \delta(c_1,c_2)(\hat{S}_i) $ can only happen when at least one of the following holds:
\begin{enumerate}
	\item $S_i$ and $\hat{S}_i$ are not from the same group.
	\item $S_i$ and $\hat{S}_i$ both from group 1 but $\hat{S}_i>c_1$ and $S_i\leq c_1$.
	\item $S_i$ and $\hat{S}_i$ both from group 1 but $\hat{S}_i\leq c_1$ and $S_i> c_1$.
	\item $S_i$ and $\hat{S}_i$ both from group 2 but $\hat{S}_i<c_2$ and $S_i\geq c_1$.
	\item $S_i$ and $\hat{S}_i$ both from group 2 but $\hat{S}_i\geq c_1$ and $S_i< c_2$.
\end{enumerate}
Since $P(\mbox{Clfdr}_i=\alpha)=P(\hlfdr_i=\alpha)=0$
The probability that $S_i$ and $\hat{S}_i$ are not from the same group is bounded by 
\begin{equation}\label{diffgp}
	P(\mbox{Clfdr}_i<\alpha,\hlfdr_i>\alpha)+ P(\mbox{Clfdr}_i>\alpha,\hlfdr_i<\alpha).
\end{equation}
Note that
\begin{align*}
	P\left(\mbox{Clfdr}_i<\alpha,\hlfdr_i>\alpha\right)&\leq P\left(\mbox{Clfdr}_i<\alpha,\hlfdr_i\in (\alpha,\alpha+\epsilon)\right)+P\left(\mbox{Clfdr}_i<\alpha,\hlfdr_i\geq \alpha+\epsilon\right)\\
	&\leq P\left(\hlfdr_i\in (\alpha,\alpha+\epsilon)\right)+P\left(\left| \mbox{Clfdr}_i-\hlfdr_i \right|>\epsilon \right).
\end{align*}
The first term on the right hand is vanishingly small as $\epsilon \rightarrow 0$ because $\hlfdr_i$ is a continuous random variable. The second term converges to $0$ by Proposition \ref{prop1}. We conclude that 
$$
P(\mbox{Clfdr}_i<\alpha,\hlfdr_i>\alpha)=o(1).
$$
Use similar argument, the remaining terms in (\ref{diffgp}) are $o(1)$, the probability of the first situation occurs is $o(1)$.

For situation 2, we have
\begin{align*}
	P\left(\hat S_i > c_1, S_i\leq  c_1 \right) &\leq P\left( S_i \leq  c_1, \hat{S}_i\in \left(c_1, c_1+ \epsilon \right) \right)+ P\left( S_i \leq c_1, \hat{S}_i\geq  + \epsilon  \right) \\
	&\leq P\left(\hat{S}_i \in \left(c_1, c_1+ \epsilon \right)\right) + P\left(\left|\hat{S}_i - S_i\right| > \epsilon \right).
\end{align*}
The first term on the right hand is vanishingly small as $\epsilon \rightarrow 0$ because $\hat{S}_i$ is a continuous random variable. The second term converges to $0$ by Lemma \ref{rconv}. we conclude that 
$$P\left(\hat S_i > c_1, S_i\leq  c_1 \right)  = o(1).$$
In a similar fashion, we can show that situation 3-5 are all $o(1)$. The lemma follows.

\subsection{Proof of Lemma \ref{rconv}}\label{pfrconv}
Let $A_\epsilon=\{ x: |\mbox{Clfdr}_i-\alpha|<\epsilon\}$. Then  $P(A_\epsilon)\rightarrow 0$ as $\epsilon\rightarrow 0$.
Let $l_0=\inf_{x_i\in A^c_\epsilon }  |\mbox{Clfdr}_i-\alpha| $ and $B_{l_0}=\{x_i: |\widehat{\mbox{Clfdr}}_i-\mbox{Clfdr}_i|>l_0/2   \}$. Since $\hlfdr_i\overset{P}{\rightarrow}\mbox{Clfdr}_i$. We have $P(B_{l_0})\rightarrow 0$. Thus $|\mbox{Clfdr}_i-\alpha|$ and $ |\hlfdr_i-\mbox{Clfdr}_i|$ are bounded below by a positive number for large $m$ except for an event that has a low probability. Note that 
$$
\hat{T}_i-T_i=\dfrac{   (\mbox{Clfdr}_i-\hlfdr_i)(x_i-\mu_0)  }{   (\hlfdr_i-\alpha)(\mbox{Clfdr}_i-\alpha)    }.
$$
It follows that $(\hat{T}_i-T_i)^2=O\left\{ (\hlfdr_i-\mbox{Lfdr}_i)^2(x_i-\mu_{0})^2  \right\}$ 
on $A^c_\epsilon \cap B^c_{l_0}$. 
Note that 
$$
(\hlfdr_i-\mbox{Lfdr}_i)^2=O_P(  \mbox{Lfdr}^2_i ). 
$$
And since $g_\mu(\cdot)$ has bounded support and the noise is Gaussian, it follows that 
$$\lim\limits_{x_i\rightarrow \pm \infty}\mbox{Lfdr}^2_i (x_i-\mu_{0})^2=0.$$
Hence $(\hat{T}_i-T_i)^2=O_P\left\{ (\hlfdr_i-\mbox{Lfdr}_i)^2 \right\}$.
Since $S_i$ and $\hat{S}_i$ are continuous function of $T_i$ and $\hat{T}_i$ respectively and $\|S_i-\hat{S}_i \|^2$ is bounded it follows that  $E\|S_i-\hat{S}_i \|^2\rightarrow 0$ and $\hat{S}_i\overset{P}{\rightarrow} S_i $.

\section{Grid Size}\label{grid}
In the proof of Proposition \ref{prop1}, we have used the fact that 
$$\mathbb{E}\|\hat{f}^m_\sigma-f_\sigma\|_2^2= O\{(mh_xh_\sigma)^{-1}+h^4_x+h^4_\sigma\}.$$
The optimal rate of $(mh_xh_\sigma)^{-1}+h^4_x+h^4_\sigma$ is $m^{-2/3}$  and is achieved when $h_x\sim h_\sigma \sim m^{-1/6}$.
Since $g_\mu$ has bounded support, for any $\mu_{j}$ we can always find $u_{i(j)}\in \{u_1,\ldots, u_k\}$ such that $|u_{i(j)}-\mu_{j}|=O(1/k)$. Let $\epsilon=|u_{i(j)}-\mu_{j}|$, then
\begin{align}\label{rateeq}
	|\phi_\tau(x-\mu_j)-\phi_\tau(x-u_{i(j)})|^2&=\dfrac{1}{2\pi\tau^2}e^{-\frac{x^2}{\tau^2}}|1-e^{\frac{2x\epsilon-\epsilon^2}{2\tau^2}}  |^2.
\end{align} 
We want the above to be of order $O(m^{-2/3})$ uniformly for any $x$. If $x$ has order greater than $\sqrt{\log m}$ then it is clear that the RHS of \eqref{rateeq} is $O(m^{-2/3})$. When $x$ has order less than $\sqrt{\log m}$, since $e^{-\frac{x^2}{\tau^2}}=O(1)$ we focus on $|1-e^{\frac{2x\epsilon-\epsilon^2}{2\tau^2}}  |^2.$ By Taylor's expansion, we have
\begin{equation*}
	|1-e^{\frac{2x\epsilon-\epsilon^2}{2\tau^2}}  |^2=O\left\{ \left(\frac{2x\epsilon-\epsilon^2}{2\tau^2}\right)^2\right\}.
\end{equation*}
It is clear that if $\epsilon=O\{1/(m^{1/3}\log m)\}$ then the above is $O(m^{-2/3})$, it follows that the a grid size of $k=O(m^{1/3}\log m)$ is sufficient.

\section{R-value, $p$-value and $q$-value}\label{sec:rv-pv-qv}

This section presents two examples that illustrate how to transform a selection procedure into an informative ranking metric using Definition 1 for the r-value. 

\begin{example}
	Suppose that our objective is to identify significant cases among multiple candidate units while controlling the per-comparison error rate (PCER). To achieve this, we can employ a simple selection rule, denoted by $I(|T_i|>\alpha)$, $i\in[m]$, where $T_i$ represents either a t-statistic or a $z$-statistic. By sequentially varying the PCER level $\alpha$ from 0 to 1, the study units can be selected in an ordered order. If we consider a scenario where the global null hypothesis is valid, and hypotheses are selected at a PCER level of $\alpha$, then the minimum $\alpha$ required for a case to be chosen is equivalent to the familiar $p$-value. The $p$-value can subsequently be used as a ranking variable to signify a unit's position in the list.
\end{example}

\begin{example}
	In the second example, let us consider the application of the adaptive $p$-value procedure \citep{benjamini2000adaptive} to select units while controlling the positive false discovery rate (pFDR) at a given level of $\alpha$. By gradually increasing $\alpha$, an informative ranking of the units can be obtained. The minimum pFDR level $\alpha$ required for a unit to be selected is known as the $q$-value \citep{Storey2002}, which can be employed as a ranking variable to indicate the unit's relative position in the list. The earlier a unit is selected, the more crucial it is deemed to be in comparison to the remaining units.
\end{example}

The r-value is a versatile concept that can be applied to a broad range of selection procedures, as illustrated by the two examples presented above. Specifically, we have shown that the $p$-value and $q$-value can be regarded as particular cases of the r-value, which can be obtained by varying the confidence level $\alpha$.

Finally, our r-value, which is based on varying $\alpha$, draws inspiration from and is closely linked to the r-value presented in \cite{HendersonNewton2014}. Nonetheless, the two definitions diverge significantly with regards to the optimization criterion and the intended goal of analysis.

\section{The Nestedness Property in Sequential Selection}\label{subsec:nest}

The topic of nested selection has been previously addressed in \cite{GuKoenker2020invidious} and \cite{HendersonNewton2014}. In an ideal scenario, if we relax the constraint by reducing $\mu_{0}$ or increasing $\alpha$, we would expect that hypotheses rejected under the stricter condition would remain rejected under the relaxed constraint. However, the oracle procedure outlined in Section \ref{oracle} may not satisfy the nestedness property defined in Definitions \ref{nest1} or \ref{nest2}. 

Section \ref{sec:r-value} introduced two notions of r-value, leading to the definition of two types of nestedness that will be discussed in the next two subsections respectively. 

\subsection{Nestedness induced by varying $\alpha$}

\begin{definition}\label{nest1} 
	Consider $\mathcal{R}^\mathcal{D}_{\alpha}$ and $\mathcal{R}^\mathcal{D}_{\alpha}$ as defined in Definition \ref{rval}. A testing procedure $\mathcal{D}$ is \emph{nested} if the rejection regions $\mathcal{R}^\mathcal{D}_{\alpha'}$ and $\mathcal{R}^\mathcal{D}_{\alpha}$ satisfy the inclusion property $\mathcal{R}^\mathcal{D}_{\alpha'}\subseteq \mathcal{R}^\mathcal{D}_{\alpha}$ for all $\alpha'<\alpha$.
\end{definition}

To illustrate why the oracle selection procedure is not nested according to Definition \ref{nest1}, consider the following example. Recall that $T_i=\dfrac{X_i-\mu_0}{\text{Clfdr}_i-\alpha}$. Suppose we have $T_1\approx T_2\approx\ldots\approx T_k$, with $T_1$ being slightly larger than $T_2,\ldots, T_k$. Additionally, assume that $\text{Clfdr}_1-\alpha>0$ and $X_1-\mu_0>0$ are both relatively large, while $\text{Clfdr}_j-\alpha>0$ and $X_j-\mu_0>0$ are relatively small for $j=2,\ldots, k$. It is worth noting that there are more than $k$ hypotheses in total, but we are focusing on these particular hypotheses for the sake of clarity and simlicity.

It is possible to select a value of $\alpha$ such that $T_1$ is rejected, while $T_2,\ldots, T_k$ are not. However, if we slightly increase the target FDR level to $\beta$, then $T_2,\ldots, T_k$ will exceed $T_1$ (assuming $\text{Clfdr}_i\geq \beta$ for $i=1,\ldots,k$). Consequently, it is possible for hypothesis 1 to be rejected at FDR level $\alpha$, but not at level $\beta$, violating the nestedness property.

\subsection{Nestedness induced by varying $\mu_0$}

\begin{definition}\label{nest2}
	Consider $\mathcal{R}^\mathcal{D}_{\mu'_{0}}$ and $\mathcal{R}^\mathcal{D}_{\mu_{0}}$ as defined in Definition \ref{rval2}. A testing procedure $\mathcal{D}$ is \textit{nested} if for all $\mu_0<\mu'_{0}$ we have $\mathcal{R}^\mathcal{D}_{\mu'_{0}}\subseteq \mathcal{R}^\mathcal{D}_{\mu_{0}}$.  
\end{definition}

To further illustrate this point, consider the following example, where the observations are generated from the following model:
$$
\mu_i\overset{iid}{\sim}U(0,10),\quad X_i|\mu_i,\sigma_i\sim N(\mu_i,\sigma^2_i).
$$
Consider the scenario in which we have two data points, $z_1=(x_1,\sigma_1)=(7.33,1)$ and $z_2=(x_2,\sigma_2)=(6.71,0.5)$. We fix $\alpha=0.1$ and set $\mu_{0}=6$. Numerical calculations reveal that both $z_1$ and $z_2$ belong to group 1, with $T_1>T_2$. Consequently, it is possible that $H_{01}$ is rejected while $H_{02}$ is not. However, if we lower $\mu_{0}$ to 5.9, $z_1$ still belongs to group 1, but $z_2$ now belongs to group 0. As a result, $H_{02}$ is rejected, but $H_{01}$ may not be. Therefore, the oracle selection procedure is not nested, as per Definition \ref{nest2}.

\subsection{Conclusion}

In summary, agreeability appears to be a more appropriate criterion for ranking procedures in the presence of heteroscedasticity. Our analysis has demonstrated that the r-values derived from Definitions 1 and 2 both meet the requirement of agreeability, which in turn results in ranking rules that are meaningful and valid.

\end{document}